\documentclass{aastex631}

\graphicspath{{./}{figures/}}

\usepackage{graphicx}

\usepackage{xcolor}

\usepackage[percent]{overpic}

\shorttitle{Tycho X-ray Ejecta Features}
\shortauthors{Millard et al.}
\begin{document}

%\includepdf[pages={12}]

\title{The 3--D X-ray Ejecta Structure of Tycho's Supernova Remnant
}

\author[0000-0001-7106-8502]{Matthew J. Millard}
\affil{Box 19059, Department of Physics, University of Texas at Arlington, Arlington, TX 76019}

\author[0000-0003-3900-7739]{Sangwook Park}
\affiliation{Box 19059, Department of Physics, University of Texas at Arlington, Arlington, TX 76019}

%\author{Jayant Bhalerao}
%\affiliation{Box 19059, Department of Physics, University of Texas at Arlington, Arlington, TX 76019}

\author[0000-0001-9267-1693]{Toshiki Sato}
\affil{Department of Physics, Rikkyo University,  3-34-1 Nishi Ikebukuro, Toshima-ku, Tokyo 171-8501, Japan}

\author[0000-0002-8816-6800]{John P. Hughes}
\affiliation{Department of Physics and Astronomy, Rutgers University, 136 Frelinghuysen Road, Piscataway, NJ 08854-8019, USA}
%\affiliation{Center for Computational Astrophysics, Flatiron Institute, 162 Fifth Avenue, New York, NY 10010, USA}

\author[0000-0002-6986-6756]{Patrick Slane}
\affiliation{Harvard-Smithsonian Center for Astrophysics, 60 Garden
Street, Cambridge, MA 02138, USA}

\author[0000-0002-7507-8115]{Daniel Patnaude}
\affiliation{Smithsonian Astrophysical Observatory, Cambridge, MA 02138, USA}

\author[0000-0003-0729-1632]{David Burrows}
\affiliation{Dept. of Astronomy \& Astrophysics, Penn State University,
University Park, PA 16802 USA}

\author[0000-0003-3494-343X]{Carles Badenes}
\affiliation{Department of Physics and Astronomy and Pittsburgh Particle
Physics, Astrophysics and Cosmology Center (PITT PACC),
University of Pittsburgh, 3941 O’Hara Street, Pittsburgh, PA
15260, USA}

%\affiliation{Institut de Ci\`{e}ncies del Cosmos (ICCUB), Universitat de
%Barcelona (IEEC-UB), Mart\'{i} Franqu\'{e}s 1, E08028 Barcelona,
%Spain}



\begin{abstract}

We present our velocity measurements of 59 clumpy, metal-rich ejecta knots in the supernova remnant (SNR) of SN 1572 (Tycho).  We use our 450 ks Chandra High Energy Transmission Grating Spectrometer observation to measure the Doppler shift of the He-like Si K$\alpha$ line-center wavelength emitted from these knots to find their line-of-sight (radial) velocities ($v_r$). We find $v_r$ up to $\sim$ 5500 km s\textsuperscript{-1}, with roughly consistent speeds between blueshifted and redshifted ejecta knots.  We also measure the proper motions (PMs) for our sample based on archival Chandra Advanced CCD Imaging Spectrometer data taken from 2003, 2009, and 2015. We estimate PMs up to 0\farcs35 yr\textsuperscript{-1}, which corresponds to a transverse velocity of about 5800 km s\textsuperscript{-1} for the distance of 3.5 kpc to Tycho. Our $v_r$ and transverse velocity measurements imply space velocities of $\sim$ 1900 -- 6000 km s\textsuperscript{-1} for the ejecta knots in Tycho. We estimate a new expansion center of R.A.(J2000) = 00\textsuperscript{h}25\textsuperscript{m}18\textsuperscript{s}.725 $\pm$ 1\textsuperscript{s}.157 and decl.(J2000) = +64$^{\circ}$08\arcmin 02\farcs5 $\pm$ 11\farcs{}2 from our PM measurements, consistent to within $\sim$ 13\arcsec{} of the geometric center.  The distribution of space velocities throughout the remnant suggests that the southeast quadrant generally expands faster than the rest of the SNR.  We find that blueshifted knots are projected more in the northern shell, while redshifted knots are more in the southern shell. The previously estimated reverse shock position is consistent with most of our estimated ejecta distribution, however some ejecta show deviations from the 1-D picture of the reverse shock.

%W

\end{abstract}

%% K

\section{Introduction} \label{sec:intro}

Type Ia supernovae (SNe) are unique among the various types of SNe because they result from a similar physical process, the thermonuclear explosion of a white dwarf that has acquired enough mass to approach the Chandrasekhar limit ($\sim$1.4 M\textsubscript{\(\odot\)}) through either a merger with or mass transfer from a binary companion.  Thus, the population of Type Ia SNe are more homogeneous than core-collapse SNe types, which have a wide range of progenitor masses.  However, the intrinsic luminosities and colors of Type Ia SNe still vary \citep{jha06,maeda11}.  Asymmetries in the explosion may contribute to their observed differences. In these cases, the viewing angle may be the cause of peculiar features observed from some SNe but not from others \citep{maeda11}. A direct method for investigating the explosion asymmetry of an SN is to measure the distribution of the ejecta in the resulting supernova remnant (SNR). The ejecta in SNRs that are only a few hundred years old have yet to be significantly disturbed by interactions with the surrounding interstellar medium (ISM). Thus, the explosion details of the SN may be preserved in the ejecta distribution. Since SNe are inherently 3--D events, the structure of an SNR should ultimately be studied in 3--D.  

%The 

Tycho's SNR (Tycho, hereafter) is the remnant of the Galactic historical supernova SN 1572. Due to its young age, its well-documented SN light curve and its close proximity to Earth ($\sim$ 2 -- 4 kpc, see \citet{hayato10}) , Tycho is an ideal choice for studying the structure of a Type Ia SNR (e.g., \citet{warren05}). Both the X-ray spectrum of the SNR \citep{badenes06} and the optical spectrum of the light echo \citep{krause08b} indicate that Tycho is the remnant of a normal Type Ia supernova, neither subluminous nor overluminous. Tycho  appears generally circular in shape, with a diameter of $\sim$ 8\arcmin{} in radio and X-ray wavelengths. XMM-Newton observations of Tycho showed an overall uniform distribution of X-ray emitting knots and filaments of shocked ejecta gas \citep{decourchelle01}. The X-ray spectrum shows bright emission from the shocked metal-rich ejecta, indicating that Si, S, Ar, Ca, and Fe abundances are several times greater than solar values \citep{hwang98}. Suzaku data showed Doppler broadening of X-ray emission lines over large areas of the SNR that suggest a generally spherical expanding ejecta shell \citep{hayato10}. \citet{williams17} measured the speeds of blueshifted and redshifted ejecta knots with the Chandra Advanced CCD Imaging Spectrometer (ACIS) data and found no clear evidence for significant asymmetry in the ejecta distribution in Tycho. In many respects, Tycho appears to be the remnant of a standard Type Ia SN explosion.  In fact, Type Ia SNe in general show a low degree of continuum polarization, implying that large deviations from spherical symmetry are not common \citep{wang08}.

Although Tycho may be regarded as the remnant of a close approximation of a standard Type Ia SN, it does contain aspherical features whose origin is not fully understood. A prominent example is in the southeast region of the SNR, where a group of metal-rich X-ray emitting ejecta clumps appears to have overtaken the forward shock \citep{vancura95,decourchelle01,wang01,fang18,williams20,sato20}.  These ejecta clumps protruding from the southeastern boundary of the SNR include Fe-rich ejecta gas that can be used to pinpoint specific nucleosynthesis models, e.g., an incomplete Si burning or an $\alpha$-rich freeze-out regime \citep{yamaguchi17}. This is in contrast to the western side of the remnant, where there is obvious separation between the ejecta and forward shock \citep{warren05}.  A Chandra study of the proper motions of reverse-shocked gas showed large azimuthal variations on the order of 50\%, while a Spitzer study suggested an ambient density enhancement by a factor of $\sim$ 3 -- 10 in the northeastern regions compared to the southwest portion of the remnant \citep{williams13}. \citet{sato19} hydrodynamically simulated Tycho's clumpy structure assuming initially clumped ejecta, as well as perfectly smooth ejecta, in all cases evolving through a uniform ambient medium. Even for the perfectly smooth case, a clumpy structure appears in the ejecta due to Rayleigh-Taylor and Kelvin-Helmholtz  instabilities.  However, the observed structure in Tycho is more consistent with an initial clumped ejecta structure from the SN rather than instabilities arising from the ejecta interaction with the ambient medium.  The optical light echo spectrum of Tycho shows an uncommon high-velocity \ion{Ca}{2} absorption feature \citep{krause08}, and thus an asymmetry in the ejecta distribution may have developed early in the evolution of the SNR, or in the SN itself. \citet{sato17tych} performed detailed spectral fits on 27 individual X-ray emitting ejecta clumps across Tycho. They found a disparity in the maximum velocities of redshifted and blueshifted features, $\lesssim$ 7800 km s \textsuperscript{-1} and $\lesssim$ 5000 km s\textsuperscript{-1}, respectively. The authors also noted large-scale He-like Si K$\alpha$ line centroid shifts across the SNR, on the order of arcminutes. They suggested that the apparent shifts may be due to differences in the intrinsic intensity of the approaching and receding sides of Tycho.

%The poin

Here, we investigate the line--of--sight velocity distributions of the clumpy metal-rich ejecta in Tycho based on our deep 450 ks Chandra High Energy Transmission Grating (HETG) observation. The high resolution HETG spectroscopy has significant advantages over the ACIS spectroscopy in several aspects. For example, the gain energies of the ACIS detectors vary by up to $0.3\%$ of the laboratory values\footnote{https://cxc.harvard.edu/proposer/POG/html/chap6.html\#tth\_sEc6.8}. For the He-like Si K$\alpha$ energy, this corresponds to an estimated uncertainty in the line--of--sight (radial) velocity up to 900 km s\textsuperscript{-1}. The type of CCD array, either ACIS-I or ACIS-S, also adds uncertainties to the line-center energies \citep{sato17tych}. ACIS data show considerable systematic uncertainties on the emission line-center energy depending on background subtraction regions. These effects contribute to overall systematic uncertainties of up to 2000 km s\textsuperscript{-1} in ACIS radial velocity ($v_r$) measurements \citep{sato17tych}. The dispersed HETG spectroscopy can avoid such systematic uncertainties associated with the ACIS spectroscopy.  Absolute wavelength uncertainties in line-center measurements with the HETG are generally $\lesssim$ 100 km s\textsuperscript{-1} \citep{marshall04,ishibashi06}\footnote{https://cxc.harvard.edu/proposer/POG/html/chap8.html}. Thus, HETG line-center energy measurements are dominated by statistical uncertainties, while the ACIS line-center measurements are dominated by systematic uncertainties.  In this work, we combine our radial velocity measurements of clumpy ejecta knots using our deep HETG data with the proper motion measurements of those knots using archival ACIS imaging data--sets to build a 3--D picture of the overall ejecta structure. In Section \ref{sec:obs}, we present the details of our deep Chandra HETG observation. In Section \ref{sec:data},  we report our analysis and results, and in Section \ref{sec:discuss} we discuss our interpretations. We conclude our study in Section \ref{sec:conclusions}.

%In Secti
\clearpage

\section{Observations} \label{sec:obs}

%\subsection{HETGS}\label{obs_hetg}

\begin{figure}
\plotone{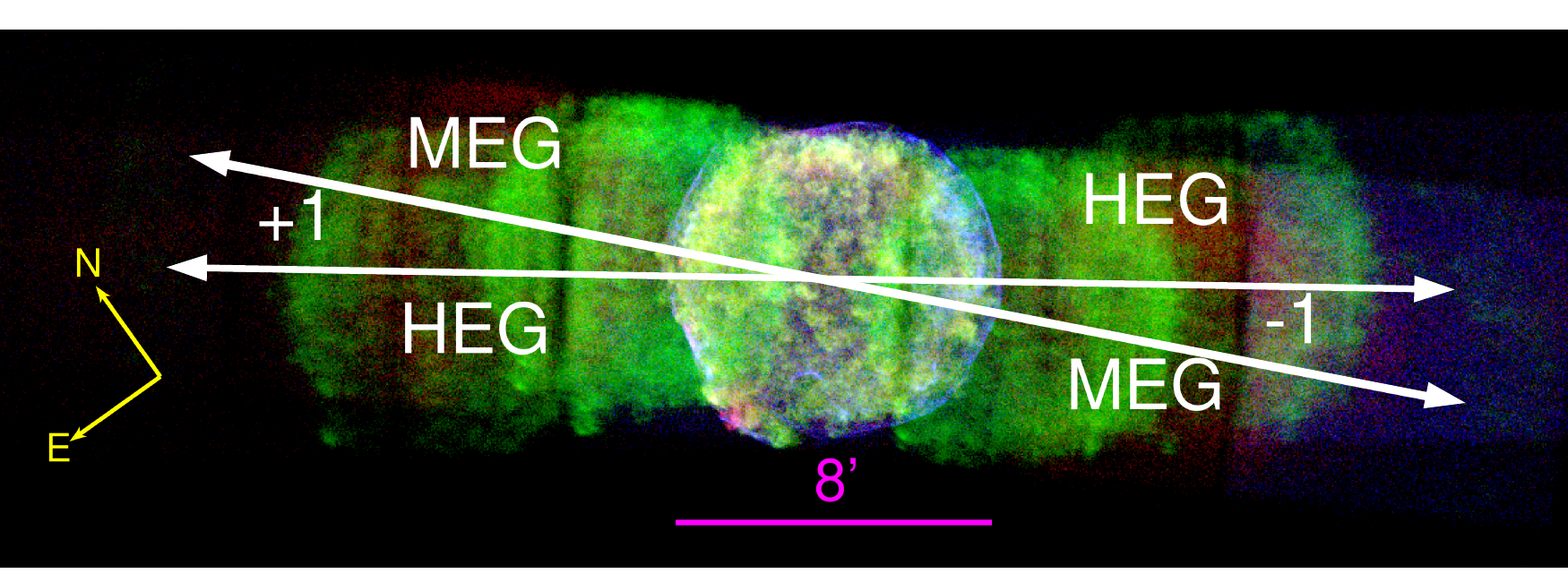}
\caption{Chandra HETG 3-color dispersed image of Tycho. Red: 0.7-1.2 keV, Green: 1.7-2.0 keV and Blue: 4.0-8.0 keV.  Our color codes are selected to represent  the Fe L line complex (red), He-like Si K$\alpha$ lines (green), and the continuum-dominated band (blue), respectively. The white arrows show the dispersion directions of the Medium and High Energy Gratings.
\label{fig:3color}}
\end{figure}

We performed our Chandra HETG observations of Tycho from 2017 October 17 to 2017 November 19. The aimpoint was set at R.A.(J2000) = 00\textsuperscript{h}25\textsuperscript{m}19\textsuperscript{s}.0, decl.(J2000) = +64$^{\circ}$08\arcmin 10\farcs{}0, which is close to the geometric center of the roughly circular SNR. The date and exposure time of each observation are listed in Table \ref{tab:obs_hetg}. The total effective exposure time is 443 ks. We processed the raw event files using Chandra Interactive Analysis of Observations (CIAO) \citep{fruscione06} version 4.11 and the Chandra Calibration Database (CALDB) version 4.8.2 to create a new level=2 event file using the CIAO command, {\tt\textbf{chandra\_repro}}.  We extracted the 1st-order dispersed spectra from a number of small regions across the SNR (Section \ref{subsec:radv}) using the TGCat scripts \citep{huenemoerder11} {\tt\textbf{tg\_create\_mask}}\footnote{http://cxc.harvard.edu/ciao/ahelp/tg\_create\_mask.html}, {\tt\textbf{tg\_resolve\_events}}\footnote{http://cxc.harvard.edu/ciao/ahelp/tg\_resolve\_events.html}, and {\tt\textbf{tgextract}}\footnote{http://cxc.harvard.edu/ciao/ahelp/tg\_extract.html}, and created the full set of corresponding detector response files using the script, {\tt\textbf{make\_responses}}, which accounts for the zeroth-order position and dispersed region size and orientation.  The HETG-dispersed image of Tycho is shown in Figure \ref{fig:3color}. 

%  generated from {\tt\textbf{tg\_create\_mask}}.

%% The "ht!" tells LaTeX to put the figure "here" first, at the "top" next
%% and to override the normal way of calculating a float position

%The 

We also use archival ACIS-I observations of Tycho (Table \ref{tab:obs_acis}) to supplement the HETG data analysis.  For the ObsIDs taken in 2009, we combined all 9 individual ObsIDs, re-projecting them onto ObsID 10095 which had the longest exposure.  The main purpose of our archival ACIS data analysis is to measure proper motions of small ejecta knots in Tycho. Thus, we re-align these ACIS images taken at three different epochs using the {\tt\textbf{reproject\_aspect}} command in CIAO, accounting for the astrometric correction based on 5 to 16 background point sources (depending on the ObsID) with their sky positions identified in the NASA/IPAC Extragalactic Database  (NED)\footnote{https://ned.ipac.caltech.edu/}.

%We shifte

\section{Data Analysis and Results} \label{sec:data}

\subsection{Region Selection} \label{subsec:hetgs}

%SNRs are 

Based on the archival Chandra ACIS observation of Tycho, we identified clumpy emission features that are bright in the Si K$\alpha$ (1.7 - 2.0 keV) band as candidate targets for HETG spectral extraction. We extracted their individual 1st-order Medium Energy Grating (MEG) spectra from our Chandra HETG observation. In Figure \ref{fig:3color}, it is clear that the zeroth-order and 1st-order images overlap. In some regions, this overlap may cause the 1st-order HETG spectrum to be contaminated by the overlapping zeroth-order emission. To avoid these regions, we compared the counts in the +1 and --1 order spectra. The majority of the knots in our sample have a +1 to --1 order counts ratio that is within a factor of $\sim$ 2. This difference in counts is consistent with the variations in effective area across the +1 and --1 order legs in the Si K$\alpha$ band\footnote{See footnote 2}, and thus their dispersed spectra are unlikely to be significantly affected by the zeroth order emission. We note that two knots in our sample (SW3 and SW7) have a larger disparity in counts between the +1 and -1 orders, up to a factor of 5. The $v_r$ estimates for these knots may be less reliable due to possible contamination from the zeroth order emission.   Applying similar methods to those developed by \citet{millard20}, we select 59 small candidate ejecta regions (angular sizes of $\sim$ 3\arcsec{} -- 10\arcsec{}) to measure the He-like Si K$\alpha$ line-center energy ($\sim$ 1.86 keV) for each region. We selected target regions such that similarly bright emission features are at least $\sim$ 20\arcsec{} away along the dispersion direction to avoid spectral contamination. We generally avoided choosing knots for our sample that are asymmetrically extended in the dispersion direction. Each region has at least 200 1st order MEG counts in the Si K$\alpha$ band.

% Regions inside the overlap area that had a similar ratio of +1 to --1 counts as those regions outside the overlap area were not considered to be significantly contaminated.

%We rejected regions where the discrepancy in counts in the He--like Si K$\alpha$ band between the +1 and --1 orders was greater than a factor of $\sim$ 5 (due to extra counts from the zeroth-order emission)

%If 

\startlongtable
\begin{deluxetable*}{ccC}
\tablecaption{Chandra HETG Observations of Tycho's SNR\label{tab:obs_hetg}}
\tablecolumns{3}
%\tablenum{1}
\tablewidth{0pt}
\tablehead{
\colhead{Observation ID} &
\colhead{Start Date} &
\colhead{Exposure Time (ks) }}
\startdata
19293	 & 	2017-10-17	 & 	49.7	\\
20813	 & 	2017-10-21	 & 	47.8	\\
20822	 & 	2017-10-23	 & 	13.9	\\
19292	 & 	2017-10-26	 & 	19.8	\\
20820	 & 	2017-10-27	 & 	30.5	\\
20819	 & 	2017-10-29	 & 	44.5	\\
19291	 & 	2017-10-30	 & 	40.0	\\
20832	 & 	2017-11-01	 & 	50.1	\\
20833	 & 	2017-11-03	 & 	34.6	\\
20834	 & 	2017-11-04	 & 	35.9	\\
20835	 & 	2017-11-06	 & 	27.6	\\
20799	 & 	2017-11-17	 & 	22.2	\\
20821	 & 	2017-11-19	 & 	25.6	\\
\enddata
%\label{tab:obs_hetg}
\end{deluxetable*}

\startlongtable
\begin{deluxetable*}{ccC}
\tablecaption{Archival Chandra ACIS-I Observations of Tycho's SNR\label{tab:obs_acis}}
\tablecolumns{3}
%\tablenum{1}
\tablewidth{0pt}
\tablehead{
\colhead{Observation ID} &
\colhead{Start Date} &
\colhead{Exposure Time (ks) }}
\startdata
3837	 & 	2003-04-29	 & 	144.6	\\
10093	 & 	2009-04-13	 & 	117.6	\\
10094	 & 	2009-04-18	 & 	89.9	\\
10095	 & 	2009-04-23 	 & 	173.4	\\
10096	 & 	2009-04-27	 & 	104.9	\\
10097	 & 	2009-04-11 	 & 	106.9	\\
10902	 & 	2009-04-15 	 & 	39.3	\\
10903	 & 	2009-04-17 	 & 	23.9	\\
10904	 & 	2009-04-13	 & 	34.7	\\
10906	 & 	2009-05-03 	 & 	40.9	\\
15998	 & 	2015-04-22	 & 	146.7	\\
\enddata
%\label{tab:obs_acis}
\end{deluxetable*}

%2003 (ObsId 3837), 2009 (ObsIds 10093, 10094, 10095, 10096, 10097, 10902, 10903, 10904, 10906), and 2015 (ObsId 15998).

\subsection{Ejecta Identification} \label{subsec:specmod}
 To identify the overabundant nature of the ejecta-dominated regions (out of our 59 selected candidate regions), we performed spectral model fits for each individual regional spectrum based on the combined 2009 archival Chandra ACIS data (merging all ObsIDs taken in 2009, to achieve the total of $\sim$ 731 ks). We fitted the observed 1.6 -- 4.5 keV band ACIS spectrum extracted from each region with an absorbed {\tt\textbf{vpshock}} model \citep{borkowski01} using the XSPEC software package version 12.10.1 \citep{arnaud96}. We estimated the background spectrum using an annulus region encircling the entire remnant. Then, we subtracted the background spectrum from the regional source spectra before fitting them with the spectral model. We fixed the absorption column at $N_H$ = 8 $\times$ 10\textsuperscript{21} cm\textsuperscript{-2} for Tycho \citep{foight16}. We allowed the electron temperature, $kT$, and ionization timescale, $\tau$ ($\tau$ = {\it n\textsubscript{e}t}, where {\it n\textsubscript{e}} is the electron density, and {\it t} is the time since being shocked), to vary. We also varied the redshift, normalization, and abundances of Si, S, Ar, and Ca. Since contributions from other elements are negligible in the 1.6 -- 4.5 keV band, we fixed all other elemental abundances at solar values \citep{wilms00}. The model gave satisfactory fits to the data, with reduced chi-squared values ranging from $\chi^{2}/dof$ = 51/73 -- 119/60. We confirm that the best--fit abundances are several times solar values, indicating that all knots in our sample are ejecta-dominated. The best-fit electron temperatures of the ejecta knots in our sample are {\it kT\textsubscript{e}} $\sim$  1 -- 5  keV, with ionization timescales $\tau$ $\sim$ 2 -- 50 $\times$ 10\textsuperscript{10} cm\textsuperscript{-3} s, generally consistent with the typical ejecta values reported in \citet{williams17}.

\subsection{Radial Velocities of Ejecta} \label{subsec:radv}

\begin{figure}
\plotone{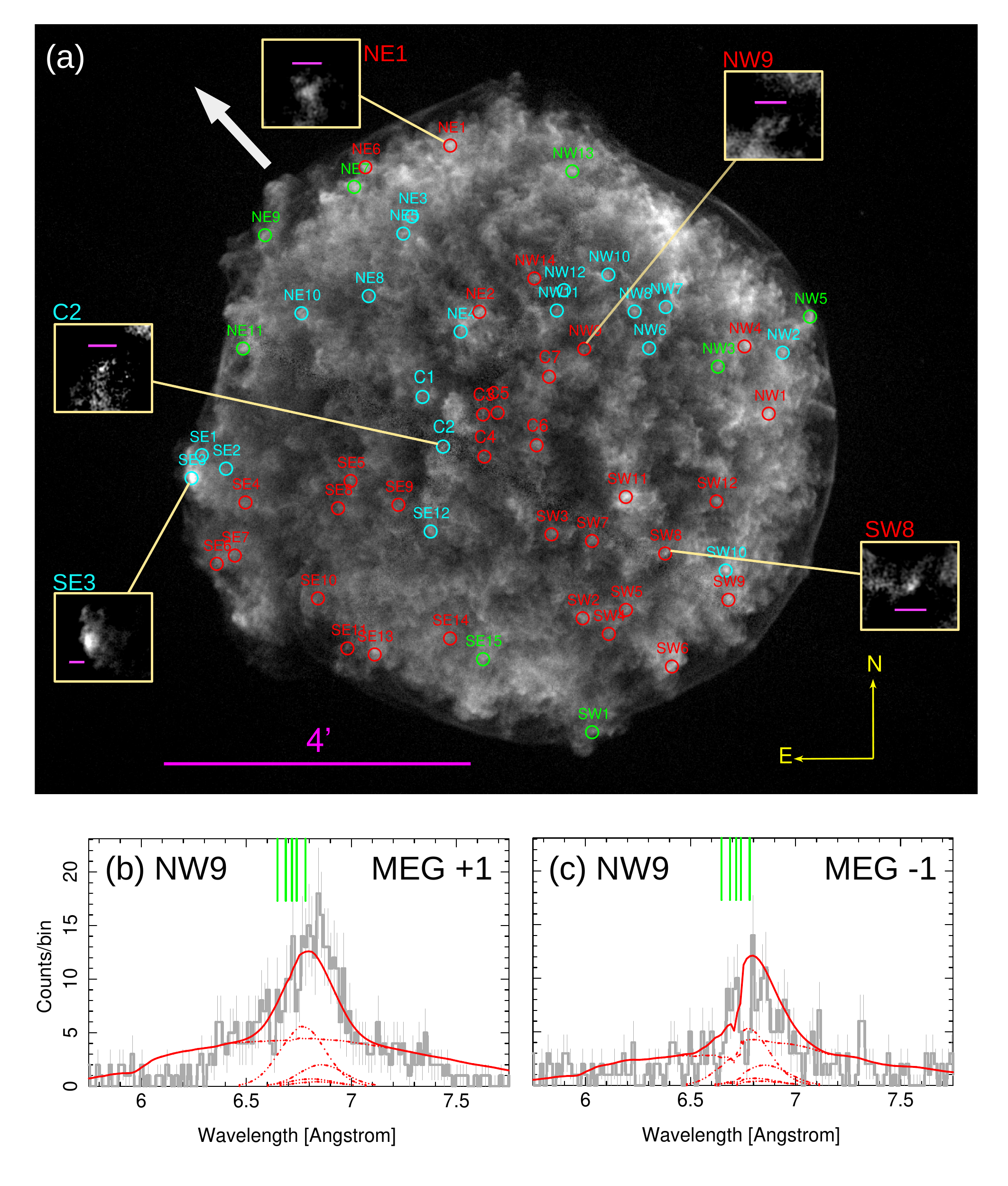}
\caption{(a): An exposure-corrected Chandra ACIS image of Tycho's SNR in the Si K$\alpha$ band (1.7 - 2.0 keV) based on the archival Chandra data taken in 2009. The fifty-nine ejecta knots analyzed in this work are marked with circles. The white arrow in the upper left indicates the dispersion direction. Cyan and red circles indicate blue- and red-shifted features, respectively, while green represents statistically negligible {\it v\textsubscript{r}} at the 90\% confidence interval. The image cutouts along the periphery show zoom-in views of example ejecta features. The scale bar in each cutout is 10\arcsec{} across. (b): An example of our line-center energy fit for region NW9.  The MEG +1 spectrum is overlaid with our best-fit Gaussian model (Gray: data; Red: model fit). The dashed lines show individual Gaussian components of our best-fit model. The vertical green lines show the locations of the rest frame He-like Si K$\alpha$ and Li-like Si XII line-center wavelengths. (c): The same as (b), however the data and model are from the MEG -1 spectrum.  
\label{fig:knotloc}}
\end{figure}

To measure the radial velocity of each X-ray emission feature in our sample, we adopt the method used in \citet{millard20}, using the Interactive Spectral Interpretation System (ISIS) software package version 1.6.2 \citep{houck00}. The spatially integrated broadband ACIS spectrum of Tycho shows bright Si, S, Ar, Ca, and Fe emission lines. However, we are interested in the HETG spectra of small features that are only a few arcseconds across. The small region sizes and the low HETG detection efficiency greatly reduces the prominence of most of these lines. Ultimately, our HETG spectra of these small emission features are dominated by the He-like Si-K$\alpha$ line. For each of our regional spectra, we measure the line-center energies in the Si K$\alpha$ band by fitting six Gaussian curves to the spectrum to account for the three He--like Si K$\alpha$ lines (6.648 \AA{} for resonance, 6.688 \AA{} for intercombination, 6.740 \AA{} for the forbidden line), and two Li-like Si XII lines at 6.717 \AA{} and 6.782 \AA{} \citep{drake88} and one for the diffuse background emission of the SNR.  We jointly fit the model to the MEG +/- 1 order spectra, tying the line-center wavelengths between spectra dispersed along the positive and negative arms. To account for the extent of the emission along the dispersion direction, the widths of the Gaussian curves are allowed to vary. The widths of the five source Gaussians are tied, while the width of the background Gaussian component varies independently and is generally much broader. Since the individual spectral lines may not be clearly resolved due to the extended nature of the ejecta knots, we fix the flux ratios among the triplet He-like Si K$\alpha$ lines and Li-like Si lines at those corresponding to the best-fit electron temperature and ionization timescale for each region (Section \ref{subsec:specmod}). We compared our measured line-center wavelength with the rest value for the resonance line at  6.648 \AA{}, which is generally the strongest among the five lines in our model.  The difference between the rest and observed values gives the Doppler shift, which we use to estimate the {\it v\textsubscript{r}} for each knot. The location of each knot is marked in Figure \ref{fig:knotloc}a, and our results are summarized in Table \ref{tab:mathmode}. Example HETG spectra and best-fit models for the +1 and -1 arms are shown in Figures \ref{fig:knotloc}b and \ref{fig:knotloc}c.

%we measure the line center energies by fitting six Gaussian curves to the spectrum - three for He--like Si K$\alpha$, two for the Si XII lines, and one for the background continuum. 

% define aliases for sato17 and williams17
\defcitealias{sato17tych}{SH17a}
\defcitealias{williams17}{W17}

\begin{figure}
\plotone{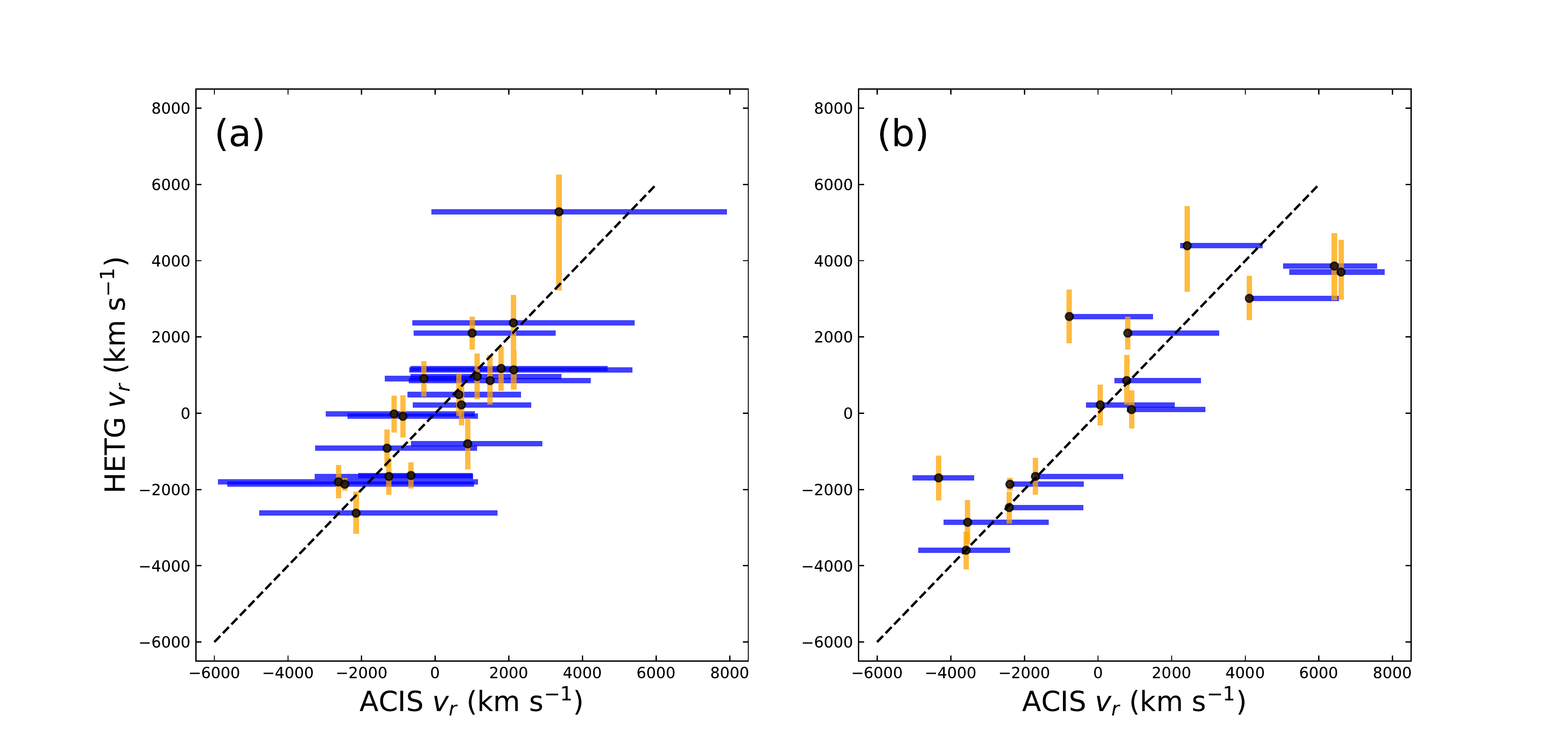}
\caption{Comparison of Chandra HETG vs. ACIS measurements of radial velocity for the common samples of ejecta knots in Tycho between this work and (a) \citetalias{williams17} and (b) \citetalias{sato17tych}. The error bars (blue) in the ACIS measurements include systematic uncertainties. 
\label{fig:acis_vs_hetg}}
\end{figure}

% Figure \ref{fig:knotloc}\textit{:Top} shows the locations of blueshifted and redshifted regions, marked by blue and red circles, respectively.
In Figure \ref{fig:knotloc}a, we show regions for which we measure Doppler shifts of Si lines. Our measured $v_r$ ranges from $\sim$ --5200 to +5300 km s\textsuperscript{-1}. We note that our sample partially overlaps with those studied by \citet{sato17tych} and \citet{williams17} (\citetalias{sato17tych} and \citetalias{williams17}, hereafter), who measured the $v_r$ of small ejecta regions in Tycho based on the lower-resolution ACIS spectroscopy: i.e., 15 and 19 regions of our sample are also included in \citetalias{sato17tych} and \citetalias{williams17}, respectively. We find general agreement between our measured values and those from ACIS data, as shown in Figure \ref{fig:acis_vs_hetg}. We found a few exceptions where our measured radial velocities are smaller than those in \citetalias{sato17tych} by a few $10^3$ km s\textsuperscript{-1} (e.g., regions C6 and SW3). The origin of the discrepancy is unclear, but may be due in part to confusion from neighboring emission. Contributions from the diffuse expanding hemispheres of the remnant may be present even in small extraction regions of only a few arcseconds in diameter, and could influence the ACIS velocity estimates. 

It is remarkable that we measure a highly significant radial velocity of $v_r$ = $-1860$ $\pm$ 170 km s\textsuperscript{-1} for the SE protrusion (region SE3 in Figure \ref{fig:knotloc}a).  This $v_r$ has been suggested based on the ACIS spectroscopy, but was not constrained due to large uncertainties of a few $10^3$ km s\textsuperscript{-1} \citepalias{sato17tych,williams17}. Based on our high resolution HETG spectroscopy, we accurately measure (within $\sim$ 10\% uncertainties) this intriguing $v_r$ for an ejecta feature projected beyond the main shell of the SNR with an order of magnitude improved accuracy.  

%The 

\startlongtable
\begin{deluxetable*}{ccCcccccccc}
\tablecaption{Radial Velocity and Proper Motion Measurements of Ejecta Features in Tycho's SNR \label{tab:mathmode}}
\tablecolumns{12}
%\tablenum{2}
\tablewidth{0pt}
\tabletypesize{\scriptsize}
\tablehead{
\colhead{Region} & % \tablenotemark{$\dagger$}
\colhead{R.A.\tablenotemark{a}} &
\colhead{Decl.\tablenotemark{a}} &
\colhead{{\it D}\tablenotemark{b}} &
\colhead{{\it v\textsubscript{r}}} &
\colhead{$\mu_{RA}$\tablenotemark{c}} & \colhead{$\mu_{Dec}$\tablenotemark{c}} & \colhead{$\mu_{Tot}$\tablenotemark{d}} & \colhead{$\eta$\tablenotemark{e}} &
\colhead{{\it v\textsubscript{s}}\tablenotemark{f}} &
 \\ 
\colhead{} & \colhead{(degree)} &
\colhead{(degree)} & \colhead{(arcmin)} & \colhead{(km s\textsuperscript{-1})} & \colhead{(arcsec yr\textsuperscript{-1})} & \colhead{(arcsec yr\textsuperscript{-1})} & \colhead{(arcsec yr\textsuperscript{-1})} & & \colhead{(km s\textsuperscript{-1})}
}
\startdata
SE1 & 6.48336 & 64.130282 & 4.07 & -2470$_{-420}^{+410}$ & -0.323 $\pm$ 0.064 & -0.002\tablenotemark{*} & 0.323 $\pm$ 0.064 & 0.6 $\pm$ 0.12 & 5910 $\pm$ 980 \\
SE2 & 6.47022 & 64.128169 & 3.74 & -670 $\pm$ 560 & -0.09 $\pm$ 0.064 & -0.057 $\pm$ 0.04 & 0.106 $\pm$ 0.066 & 0.21 $\pm$ 0.13 & 1890 $\pm$ 1040 \\
SE3 & 6.49035 & 64.125542 & 4.28 & -1860 $\pm$ 170 & -0.299 $\pm$ 0.063 & -0.096 $\pm$ 0.04 & 0.314 $\pm$ 0.064 & 0.56 $\pm$ 0.11 & 5530 $\pm$ 1000 \\
SE4 & 6.46192 & 64.12032 & 3.6 & 1650$_{-710}^{+690}$ & - & - & - & - & - \\
SE5 & 6.40962 & 64.125008 & 2.2 & 3840 $\pm$ 820 & - & - & - & - & - \\
SE6 & 6.47617 & 64.106979 & 4.21 & 1660 $\pm$ 660 & -0.249 $\pm$ 0.062 & -0.061 $\pm$ 0.038 & 0.257 $\pm$ 0.063 & 0.45 $\pm$ 0.11 & 4580 $\pm$ 1000 \\
SE7 & 6.46718 & 64.10879 & 3.95 & 1380$_{-620}^{+610}$ & -0.314 $\pm$ 0.063 & 0.06 $\pm$ 0.035 & 0.32 $\pm$ 0.063 & 0.6 $\pm$ 0.12 & 5490 $\pm$ 1020 \\
SE8 & 6.41552 & 64.118707 & 2.47 & 2780$_{-710}^{+690}$ & -0.238 $\pm$ 0.061 & -0.086 $\pm$ 0.033 & 0.253 $\pm$ 0.061 & 0.76 $\pm$ 0.18 & 5040 $\pm$ 930 \\
SE9 & 6.3844 & 64.120199 & 1.69 & 4210$_{-1030}^{+980}$ & - & - & - & - & - \\
SE10 & 6.42579 & 64.099514 & 3.29 & 2780 $\pm$ 680 & - & - & - & - & - \\
SE11 & 6.4111 & 64.088671 & 3.49 & 1070$_{-740}^{+730}$ & - & - & - & - & - \\
SE12 & 6.36942 & 64.113216 & 1.65 & -920 $\pm$ 490 & - & - & - & - & - \\
SE13 & 6.39798 & 64.086552 & 3.39 & 2100 $\pm$ 430 & - & - & - & - & - \\
SE14 & 6.36005 & 64.090842 & 2.72 & 5180$_{-740}^{+770}$ & -0.012\tablenotemark{*} & -0.171 $\pm$ 0.04 & 0.172 $\pm$ 0.07 & 0.44 $\pm$ 0.19 & 5920$_{-860}^{+880}$ \\
SE15 & 6.34372 & 64.085777 & 2.93 & -20\tablenotemark{*} & -0.053\tablenotemark{*} & -0.326 $\pm$ 0.037 & 0.33 $\pm$ 0.067 & 0.8 $\pm$ 0.17 & 5480 $\pm$ 1110  \\
NE1 & 6.36031 & 64.198296 & 3.95 & 960 $\pm$ 600 & 0.001\tablenotemark{*} & 0.354 $\pm$ 0.038 & 0.354 $\pm$ 0.068 & 0.71 $\pm$ 0.13 & 5950 $\pm$ 1120 \\
NE2 & 6.34541 & 64.162038 & 1.74 & 2460$_{-1070}^{+1010}$ & -0.001\tablenotemark{*} & 0.158 $\pm$ 0.041 & 0.158 $\pm$ 0.071 & 0.77 $\pm$ 0.3 & 3600$_{-1130}^{+1100}$ \\
NE3 & 6.37907 & 64.182406 & 3.19 & -2290$_{-600}^{+590}$ & -0.029\tablenotemark{*}& 0.243 $\pm$ 0.038 & 0.245 $\pm$ 0.068 & 0.61 $\pm$ 0.16 & 4670$_{-1030}^{+1020}$ \\
NE4 & 6.35505 & 64.156519 & 1.52 & -3590$_{-500}^{+490}$ & -0.062\tablenotemark{*} & 0.077 $\pm$ 0.038 & 0.099 $\pm$ 0.067 & 0.54 $\pm$ 0.33 & 3950$_{-650}^{+640}$ \\
NE5 & 6.38373 & 64.178972 & 3.06 & -4210$_{-620}^{+600}$ & - & - & - & - & - \\
NE6 & 6.40231 & 64.193381 & 4.06 & 910$_{-470}^{+460}$ & -0.125 $\pm$ 0.063 & 0.195 $\pm$ 0.038 & 0.231 $\pm$ 0.067 & 0.45 $\pm$ 0.12 & 3940 $\pm$ 1090 \\
NE7 & 6.40808 & 64.189185 & 3.92 & -400\tablenotemark{*} & -0.16 $\pm$ 0.063 & 0.207 $\pm$ 0.04 & 0.262 $\pm$ 0.068 & 0.53 $\pm$ 0.13 & 4360 $\pm$ 1130 \\
NE8 & 6.40062 & 64.165145 & 2.66 & -1800 $\pm$ 440 & -0.126 $\pm$ 0.061 & 0.158 $\pm$ 0.034 & 0.202 $\pm$ 0.063 & 0.61 $\pm$ 0.18 & 3800$_{-940}^{+950}$ \\
NE9 & 6.45171 & 64.179368 & 4.23 & 490\tablenotemark{*} & -0.133 $\pm$ 0.062 & 0.159 $\pm$ 0.035 & 0.207 $\pm$ 0.064 & 0.38 $\pm$ 0.11 & 3460 $\pm$ 1050 \\
NE10 & 6.43524 & 64.161668 & 3.26 & -1660$_{-480}^{+490}$ & -0.204 $\pm$ 0.062 & 0.049 $\pm$ 0.035 & 0.21 $\pm$ 0.062 & 0.51 $\pm$ 0.14 & 3860 $\pm$ 950 \\
NE11 & 6.46282 & 64.153989 & 3.72 & 100\tablenotemark{*} & -0.173 $\pm$ 0.062 & 0.021\tablenotemark{*} & 0.174 $\pm$ 0.062 & 0.36 $\pm$ 0.12 & 2900 $\pm$ 1030 \\
NW1 & 6.20057 & 64.139647 & 3.35 & 1900 $\pm$ 570 & 0.231 $\pm$ 0.064 & -0.077 $\pm$ 0.038 & 0.244 $\pm$ 0.064 & 0.53 $\pm$ 0.14 & 4470 $\pm$ 990 \\
NW2 & 6.1934 & 64.153122 & 3.7 & -800 $\pm$ 670 & 0.264 $\pm$ 0.064 & 0.103 $\pm$ 0.039 & 0.284 $\pm$ 0.065 & 0.57 $\pm$ 0.13 & 4780 $\pm$ 1070 \\
NW3 & 6.22578 & 64.149849 & 2.84 & -310\tablenotemark{*} & 0.17 $\pm$ 0.064 & 0.071 $\pm$ 0.036 & 0.185 $\pm$ 0.064 & 0.48 $\pm$ 0.17 & 3080 $\pm$ 1060 \\
NW4 & 6.21326 & 64.154398 & 3.24 & 1180 $\pm$ 580 & 0.227 $\pm$ 0.064 & 0.083 $\pm$ 0.038 & 0.242 $\pm$ 0.065 & 0.56 $\pm$ 0.15 & 4190 $\pm$ 1050 \\
NW5 & 6.17947 & 64.160641 & 4.2 & 220\tablenotemark{*} & 0.289 $\pm$ 0.065 & 0.155 $\pm$ 0.037 & 0.328 $\pm$ 0.066 & 0.6 $\pm$ 0.12 & 5450 $\pm$ 1100 \\
NW6 & 6.25986 & 64.154143 & 2.15 & -3980$_{-660}^{+650}$ & 0.162 $\pm$ 0.063 & 0.039 $\pm$ 0.037 & 0.167 $\pm$ 0.064 & 0.6 $\pm$ 0.22 & 4850$_{-820}^{+810}$ \\
NW7 & 6.25252 & 64.162735 & 2.62 & -1000 $\pm$ 600 & 0.126 $\pm$ 0.064 & 0.14 $\pm$ 0.04 & 0.189 $\pm$ 0.068 & 0.55 $\pm$ 0.19 & 3290 $\pm$ 1090 \\
NW8 & 6.26811 & 64.162003 & 2.3 & -3430 $\pm$ 500 & 0.116 $\pm$ 0.063 & 0.081 $\pm$ 0.037 & 0.141 $\pm$ 0.065 & 0.47 $\pm$ 0.21 & 4150$_{-730}^{+740}$ \\
NW9 & 6.29317 & 64.153676 & 1.49 & 5320$_{-950}^{+920}$ & - & - & - & - & - \\
NW10 & 6.2808 & 64.16993 & 2.48 & -2620$_{-540}^{+560}$ & 0.029\tablenotemark{*} & 0.102 $\pm$ 0.034 & 0.106 $\pm$ 0.064 & 0.34 $\pm$ 0.19 & 3160$_{-740}^{+750}$ \\
NW11 & 6.30679 & 64.162212 & 1.78 & -1120 $\pm$ 640 & - & - & - & - & - \\
NW12 & 6.30327 & 64.166397 & 2.05 & -5220 $\pm$ 890 & - & - & - & - & - \\
NW13 & 6.29901 & 64.192441 & 3.58 & -80* & -0.030\tablenotemark{*} & 0.147 $\pm$ 0.04 & 0.15 $\pm$ 0.07 & 0.33 $\pm$ 0.14 & 2490 $\pm$ 1160 \\
NW14 & 6.31804 & 64.168952 & 2.11 & 5280$_{-2070}^{+980}$ & 0.031\tablenotemark{*}& 0.149 $\pm$ 0.036 & 0.152 $\pm$ 0.065 & 0.58 $\pm$ 0.23 & 5860$_{-1920}^{+1000}$ \\
SW1 & 6.28838 & 64.07014 & 3.97 & 160\tablenotemark{*} & - & - & - & - & - \\
SW2 & 6.29334 & 64.094728 & 2.53 & 1140 $\pm$ 520 & 0.076 $\pm$ 0.063 & -0.194 $\pm$ 0.039 & 0.208 $\pm$ 0.068 & 0.57 $\pm$ 0.2 & 3640 $\pm$ 1080 \\
SW3 & 6.30962 & 64.113309 & 1.33 & 3860$_{-900}^{+870}$$^{\dagger{}}$ & - & - & - & - & - \\
SW4 & 6.2801 & 64.091581 & 2.84 & 2060$_{-560}^{+580}$ & 0.061\tablenotemark{*} & -0.214 $\pm$ 0.036 & 0.223 $\pm$ 0.065 & 0.55 $\pm$ 0.17 & 4240 $\pm$ 980 \\
%SW5 & 6.29859 & 64.114338 & 1.41 & 3850$_{-800}^{+840}$ & - & - & - & - & - \\
SW5 & 6.27296 & 64.096338 & 2.68 & 840$_{-480}^{+470}$ & 0.111 $\pm$ 0.063 & -0.106 $\pm$ 0.035 & 0.153 $\pm$ 0.064 & 0.4 $\pm$ 0.18 & 2680 $\pm$ 1020 \\
SW6 & 6.24856 & 64.084384 & 3.64 & 860$_{-630}^{+680}$ & 0.021\tablenotemark{*} & -0.208 $\pm$ 0.035 & 0.209 $\pm$ 0.065 & 0.41 $\pm$ 0.13 & 3570 $\pm$ 1060 \\
SW7 & 6.28942 & 64.111983 & 1.67 & 3010$_{-580}^{+590}$$^{\dagger{}}$ & - & - & - & - & - \\
SW8 & 6.25281 & 64.108776 & 2.49 & 2370$_{-720}^{+730}$ & 0.018\tablenotemark{*} & -0.073 $\pm$ 0.038 & 0.076 $\pm$ 0.067 & 0.21 $\pm$ 0.2 & 2680$_{-820}^{+830}$ \\
SW9 & 6.22021 & 64.099009 & 3.52 & 2540 $\pm$ 700 & 0.153 $\pm$ 0.063 & -0.068 $\pm$ 0.037 & 0.168 $\pm$ 0.064 & 0.34 $\pm$ 0.13 & 3760 $\pm$ 920 \\
SW10 & 6.22228 & 64.105266 & 3.26 & -1630 $\pm$ 340 & 0.21 $\pm$ 0.063 & -0.154 $\pm$ 0.038 & 0.26 $\pm$ 0.065 & 0.57 $\pm$ 0.15 & 4620 $\pm$ 1020 \\
SW11 & 6.27209 & 64.12145 & 1.65 & 2830$_{-680}^{+700}$ & - & - & - & - & - \\
SW12 & 6.22676 & 64.120373 & 2.78 & 740$_{-630}^{+640}$ & 0.118 $\pm$ 0.063 & -0.081 $\pm$ 0.036 & 0.143 $\pm$ 0.064 & 0.37 $\pm$ 0.17 & 2490 $\pm$ 1030 \\
C1 & 6.3738 & 64.143253 & 1.32 & -1700 $\pm$ 590 & - & - & - & - & - \\
C2 & 6.36353 & 64.132469 & 0.93 & -2860$_{-570}^{+580}$ & - & - & - & - & - \\
C3 & 6.34361 & 64.139439 & 0.52 & 4460$_{-770}^{+780}$ & - & - & - & - & - \\
C4 & 6.34296 & 64.130289 & 0.45 & 4060 $\pm$ 880 & - & - & - & - & - \\
C5 & 6.33638 & 64.13981 & 0.41 & 3680$_{-950}^{+870}$ & - & - & - & - & - \\
C6 & 6.31697 & 64.132757 & 0.3 & 3700$_{-730}^{+850}$ & - & - & - & - & - \\
C7 & 6.31072 & 64.14762 & 0.93 & 4390$_{-1210}^{+1040}$ & - & - & - & - & - \\
\enddata
\tablenotetext{}{Errors represent a 90\% confidence interval unless otherwise noted.}
\tablenotetext{a}{Position in 2015 (J2000).}
\tablenotetext{b}{Projected angular distance from our estimated kinematic center (see Section \ref{sec:rev_shoc_center})}
\tablenotetext{c}{Includes systematic uncertainties (see Section \ref{subsec:propmot}).}
%; R.A.(J2000) = 17\textsuperscript{h} 30\textsuperscript{m} 41\textsuperscript{s}.321 and Declination(J2000) = -21$^{\circ}$ 29\arcmin{} 30\arcsec.51, with uncertainties of $\sigma_{R.A.}= \pm$  0.073\arcmin{} and $\sigma_{Dec}= \pm$ 0.072\arcmin{}, respectively. }
\tablenotetext{d}{$\mu_{Tot} = \sqrt[]{\mu_{R.A.}^{2}+\mu_{decl.}^{2}}$.}
\tablenotetext{e}{Expansion index (see Section \ref{subsec:propmot}).}
\tablenotetext{f}{Estimated space velocity for a distance of 3.5 kpc.}
\tablenotetext{*}{The error interval includes 0, and thus the direction of motion is uncertain.  We show only our best-fit value.}
\tablenotetext{\dagger}{This estimate may be affected due to spectral contamination from zeroth order emission (Section \ref{subsec:hetgs}).}
\label{table:all}
\end{deluxetable*}

%\clearpage

%\clearpage

\subsection{Ejecta Proper Motions} \label{subsec:propmot}

Based on the archival Chandra ACIS-I data from 2003, 2009, and 2015 (Table \ref{tab:obs_acis}), we estimate the proper motions of the ejecta regions in our sample. To measure the proper motions, we apply the methods described in \citet{sato18}. To find the position of each knot at different epochs, we took the image from the long observation in 2009 (ObsID 10095) as the reference image and compared it to the images from the 2003 and 2015 epochs filtered to the 1.6 -- 4.5 keV band, which is dominated by the Si K$\alpha$ line. We incrementally shifted the 2003 and 2015 images in R.A. and decl. until a statistically good match with the reference image was obtained, i.e., the Cash statistic \citep{cash79} was minimized. To estimate the systematic uncertainties, we applied this image fitting method to five background point sources. We find the systematic uncertainties of our method to be $\sigma_{\mu_{RA}}= 0\farcs{}06$ yr\textsuperscript{-1} and $\sigma_{\mu_{decl.}}= 0\farcs{}03$ yr\textsuperscript{-1}, in reasonable agreement with the uncertainties estimated in \citet{katsuda10}. We were able to successfully measure proper motions for 37 of the 59 knots in our sample. For other knots, it was difficult to measure proper motions because they were faint or contaminated by complex emission features in the immediate surroundings. Regions projected close to the center of the SNR do not show measurable proper motions (as perhaps expected), and thus their space velocity is dominated by their radial velocity.

%% The "ht!" tells LaTeX to put the figure "here" first, at the "top" next
%% and to override the normal way of calculating a float position

% Plot comparing ACIS vs HETG radial velocities
%%%%%%%%%%%%%%%%%%%%%%%%%%%%%%%%%%%%%%%%%%%%%%%%
%%%%%%%%%%%%%%%%%%%%%%%%%%%%%%%%%%%%%%%%%%%%%%%%

%Ejecta

% Show image of PM measurements using arrows
%%%%%%%%%%%%%%%%%%%%%%%%%%%%%%%%%%%%%%%%%%%%%%%%
\begin{figure}
\plotone{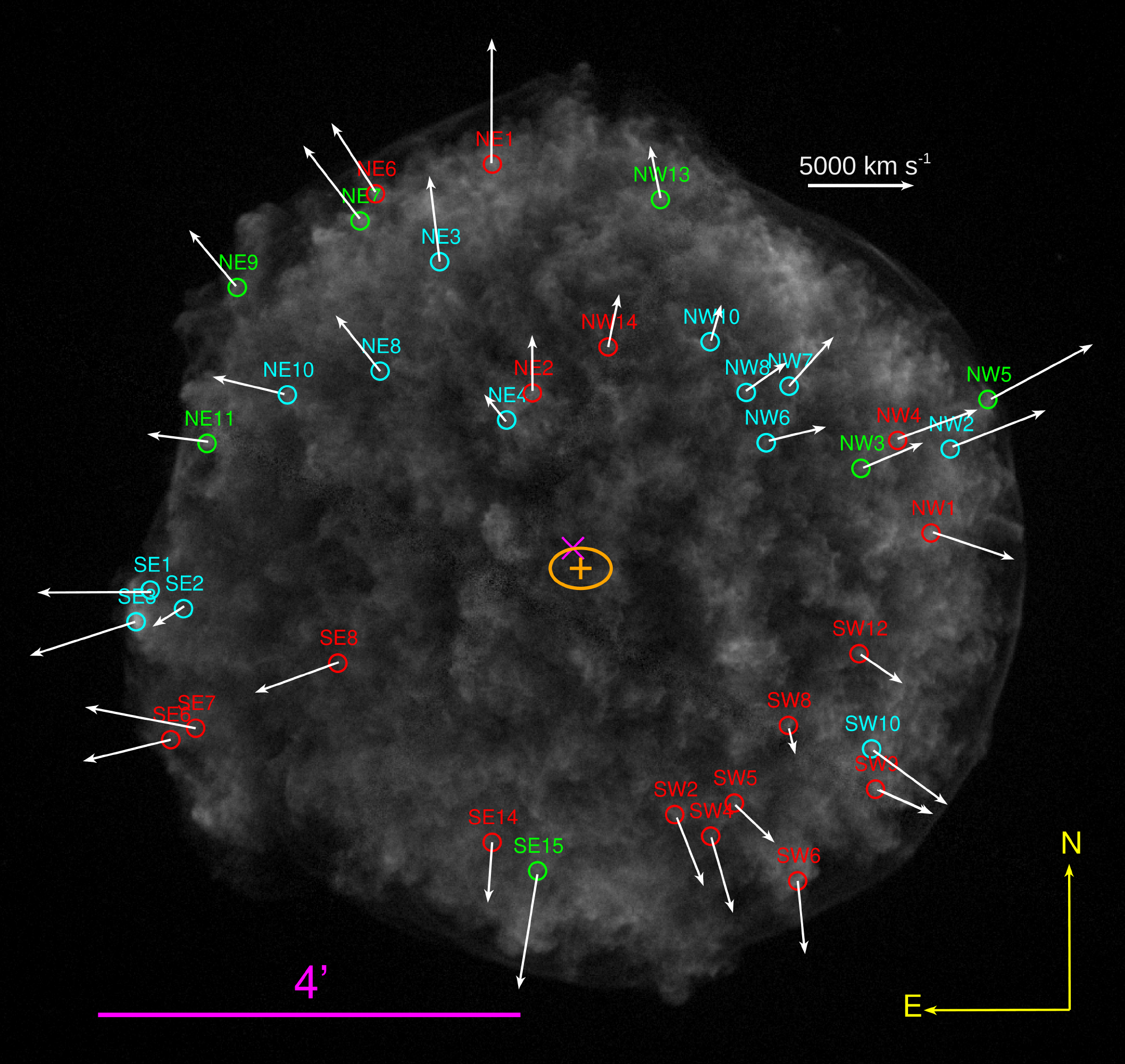}
\caption{The subset of our regions where we measure the proper motion. Each white arrow shows the direction and relative magnitude of the proper motion for each knot. The length of the white arrow at the top right indicates a speed of 5000 km s\textsuperscript{-1}.  The orange cross and ellipse indicate the position and uncertainty of our estimated kinematic center based on our proper motion measurements of ejecta knots (see Section \ref{sec:rev_shoc_center}). The magenta ``X'' indicates the geometric center \citep{warren05}.  The image is the same as in Figure \ref{fig:knotloc}, scaled to make the arrows more visible.
\label{fig:pm_arrows}}
\end{figure}
%%%%%%%%%%%%%%%%%%%%%%%%%%%%%%%%%%%%%%%%%%%%%%%%

The results of our proper motion measurements are summarized in Table \ref{tab:mathmode}. Our measured values range from $-$0\farcs{}32 yr\textsuperscript{-1} to +0\farcs{}29 yr\textsuperscript{-1} in R.A. and $-$0\farcs{}33 yr\textsuperscript{-1} to +0\farcs{}35 yr\textsuperscript{-1} in decl. We define the expansion index, $\eta$, as $\mu_{Tot}/(D/t_{age})$, where $t_{age}$ is the age of Tycho (445 years in this study), and $D$ is the estimated angular distance from the geometric center.  Our proper motion measurements suggest that all of the knots in the sample have undergone some significant deceleration, ranging from $\eta$ = 0.21 to 0.80, with an average $\eta = 0.51$. The proper motion directions are shown in Figure \ref{fig:pm_arrows}.

We combine the radial velocity and proper motion measurements to estimate the 3--D space velocity of regions in our sample. We adopt a distance of 3.5 kpc to Tycho \citep{williams13}. At this distance, the transverse velocities of knots at the boundary and the radial velocities of knots projected near the center of the SNR generally agree, $v_r \sim 5500$ km s\textsuperscript{-1}, which is consistent with the maximum range of space velocities for ejecta regions in our sample.  Combining the radial velocity and proper motions we estimate the space velocities of $\sim$ 1900 -- 6000 km s\textsuperscript{-1}, with an average $v_s \sim$ 4200 km s\textsuperscript{-1}. These velocity ranges are in plausible agreement with those estimated by \citetalias{sato17tych} and \citetalias{williams17}, but with uncertainties smaller by a factor of $\sim$ 3 on average.

% The average 

%Figure \ref{fig:deltac} shows best-fit locations of the knots, demonstrating their positional changes over 12 years. 

%\newpage

\section{Discussion} \label{sec:discuss}

\subsection{Azimuthal Variations in Ejecta Velocity}
\begin{figure}
\plotone{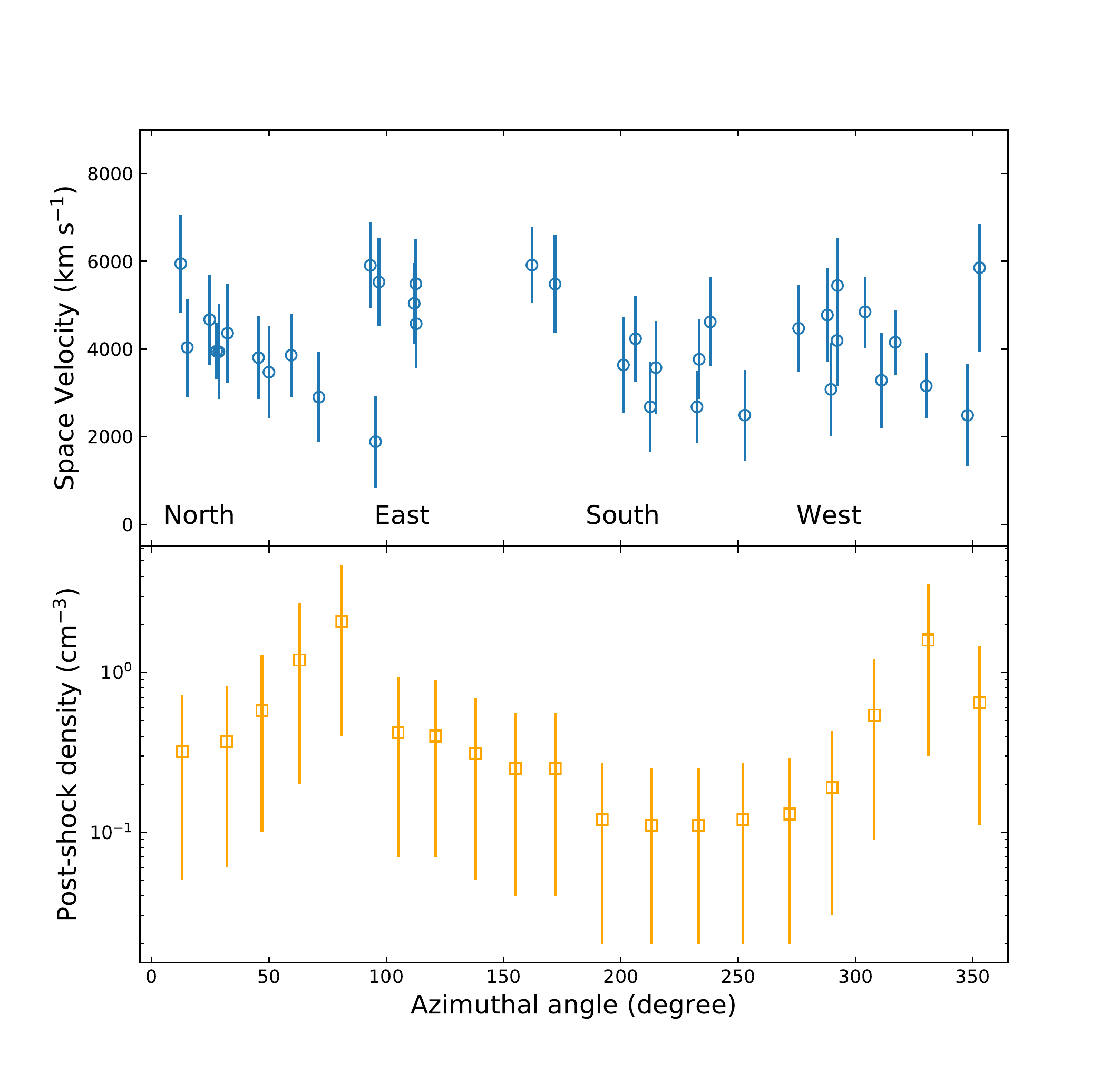}
\caption{The upper panel shows the azimuthal distribution of our estimated ejecta space velocities in Tycho. The bottom panel shows the estimated post-shock densities along the rim from \citet{williams13}.
\label{fig:space_vel_az_angle}}
\end{figure}

In Figure \ref{fig:space_vel_az_angle}, we plot our estimated space velocity for each knot against its azimuthal angle (position angles measured counter-clockwise from north). For knots projected closer to the center of the remnant, their true location along the periphery of the SNR is more uncertain. Thus, we only included knots with projected positions offset from the center where we have firmly estimated their proper motions. The ejecta knots in the southeast (SE) quadrant of Tycho have $v_s \sim$ 6000 km s\textsuperscript{-1}, and thus appear to be among the fastest-moving knots ($\sim$ 40\% faster than the average space velocity of our sample). This high space velocity is in plausible agreement with the presence of ejecta bullet-like features (protrusions extending beyond the main SNR shell, \citep{wang01}) where high-speed overdense clumps overtake the forward shock. However, the protruding knots are not individual ejecta features like in \citet{wang01}, but are part of a large-scale portion of the ejecta that was propelled from the explosion more energetically than elsewhere in the remnant. The densities along the SE rim are larger by a factor of a few than those in the southwest (SW) \citep{williams13}. However, the ejecta space velocities in the SE are faster by a factor of $\sim 2$ than those in the SW. Thus, the ejecta velocities in SE regions may not be directly related to a rarefied ISM in that direction, but probably due to their intrinsically energetic nature. It is interesting to note that there is a prominent high-speed ejecta knot (NW5) approaching $v_{space} \sim$ 6000 km s\textsuperscript{-1} projected at the northwest (NW) boundary, in a nearly opposite direction from the protruding SE knots. While it is tempting to speculate strong ejecta outflows along the SE-NW axis connecting these particularly fast-moving knots, we find no additional substantial evidence to support such a bi-polar ejecta outflow along this axis.

In the northeast (NE), from position angles $10\degr{}$ to 100$\degr{}$, the ejecta space velocities appear to decrease, from $\sim$ 6000 km s\textsuperscript{-1} to $\sim$ 2000 km s\textsuperscript{-1}.  The space velocity then sharply rises back up to $\sim$ 6000 km s\textsuperscript{-1} for ejecta knots in the SE from position angles $\sim$ 100\degr{} -- 170\degr{}, before decreasing again to 2000--4000 km s\textsuperscript{-1} in the SW from 200\degr{} to 250\degr{}.  Some decreases in ejecta velocity with azimuthal angle are coincident with increasing ambient density, suggesting an origin from the SNR's interaction with a dense surrounding medium.  A Spitzer study of the ratio of the 70 to 24 $\mu$m fluxes in Tycho revealed an increase of post-shock densities at the rim from azimuthal angles of roughly 10 -- 80\degr{} and 300 -- 330\degr{} \citep{williams13}, similar to the angle ranges of decreasing velocity (see Figure \ref{fig:space_vel_az_angle}). The ejecta in these regions may have been slowed either by direct interaction with the higher-density ISM gas or by an enhanced reverse shock that developed due to the shock-ISM interaction, or a combination of both.

%In 

% Plot of space velocity vs azimuthal angle
%%%%%%%%%%%%%%%%%%%%%%%%%%%%%%%%%%%%%%%%%%%%%%%%
%%%%%%%%%%%%%%%%%%%%%%%%%%%%%%%%%%%%%%%%%%%%%%%%

%\newpage

\subsection{3--D Ejecta Structure}\label{sec:3d_ejec_struc}
\begin{figure}
%\plottwo{test_3D_03-06-2022_into_z-axis.pdf}{test_3D_03-06-2022_into_x-axis.pdf}
\gridline{\fig{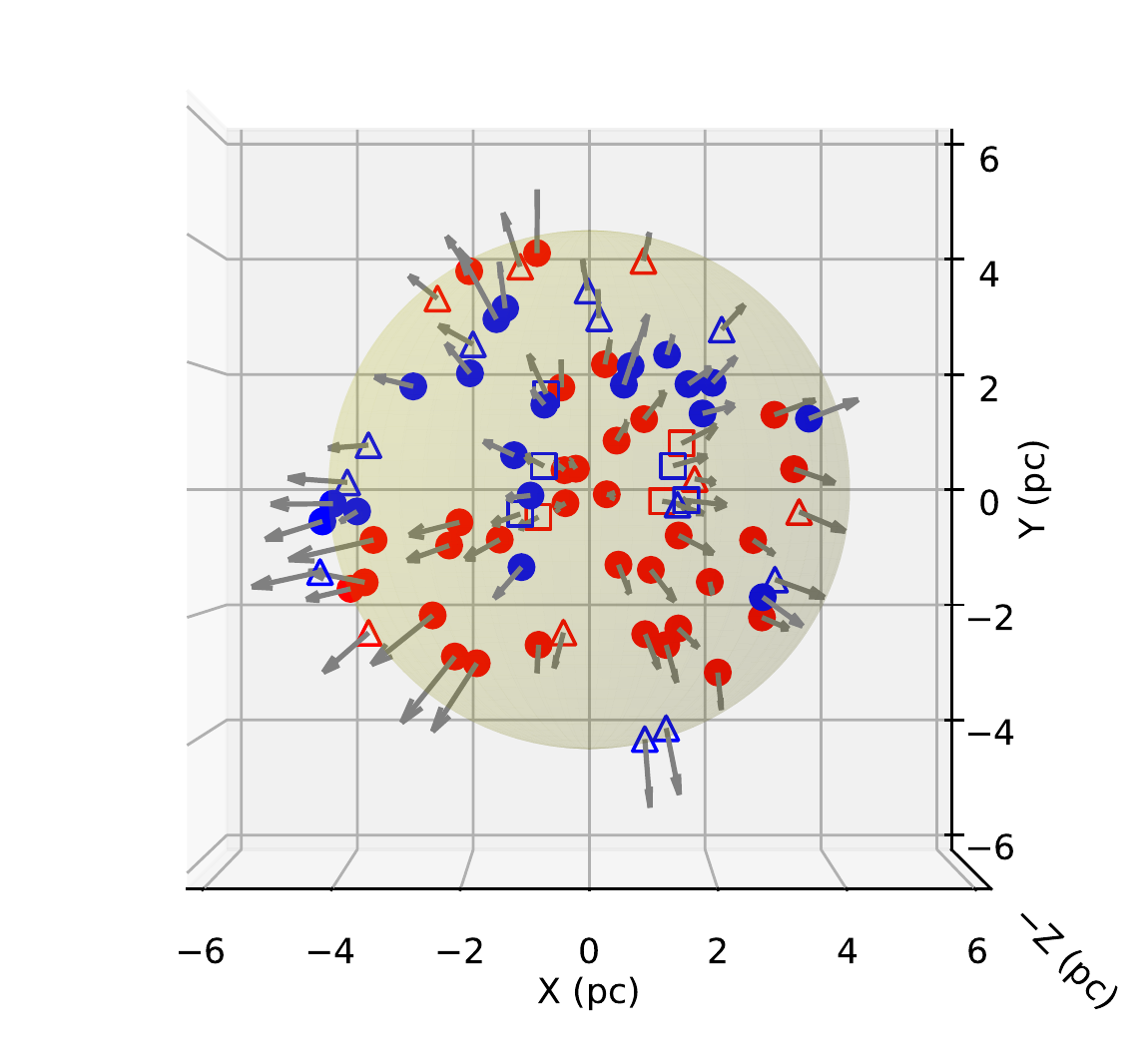}{0.49\textwidth}{(a)}
          \fig{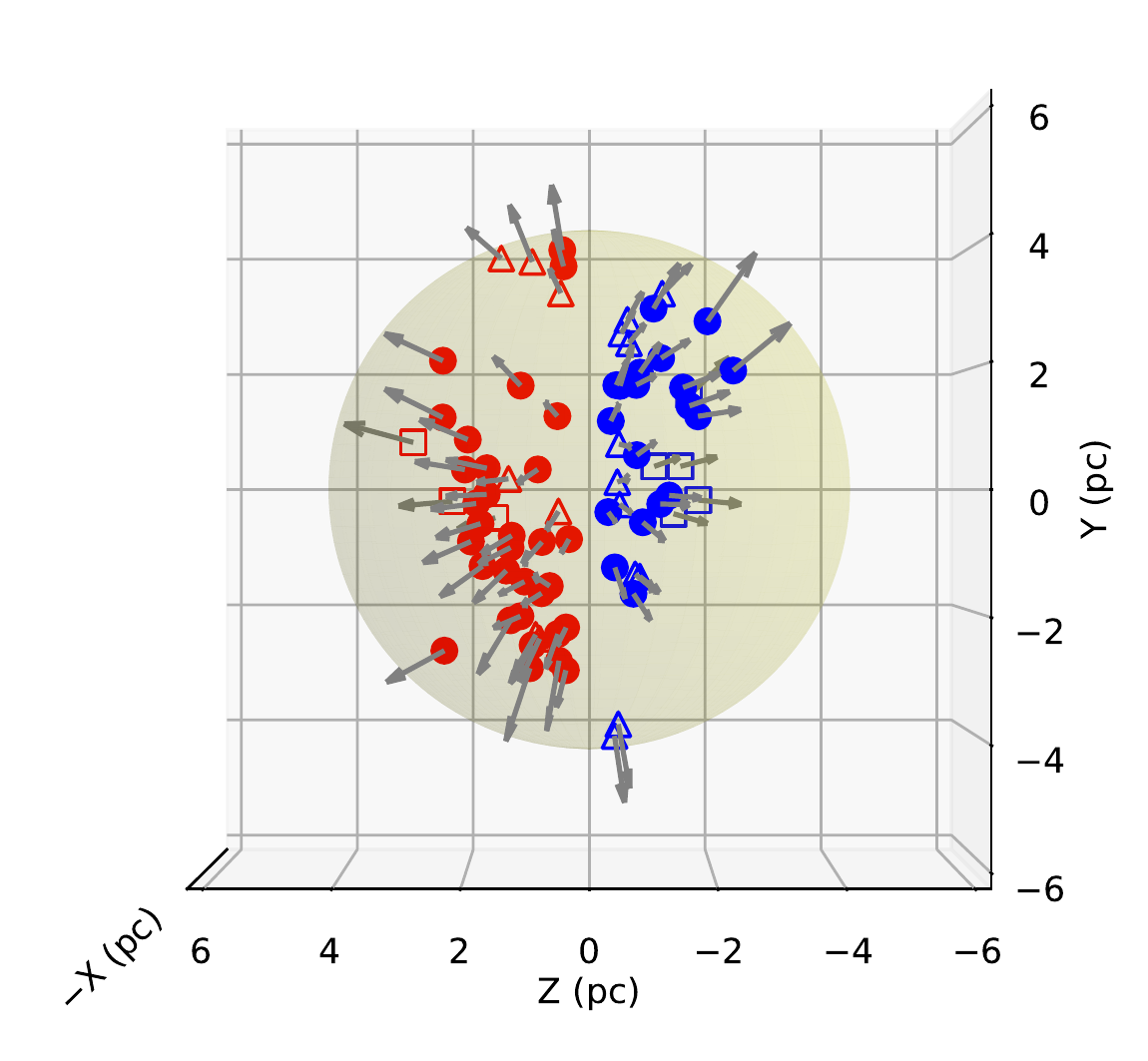}{0.49\textwidth}{(b)}
          %\fig{test_3D_05-31-2022_x-direction.pdf}{0.49\textwidth}{(c)}
          }
\gridline{
          \fig{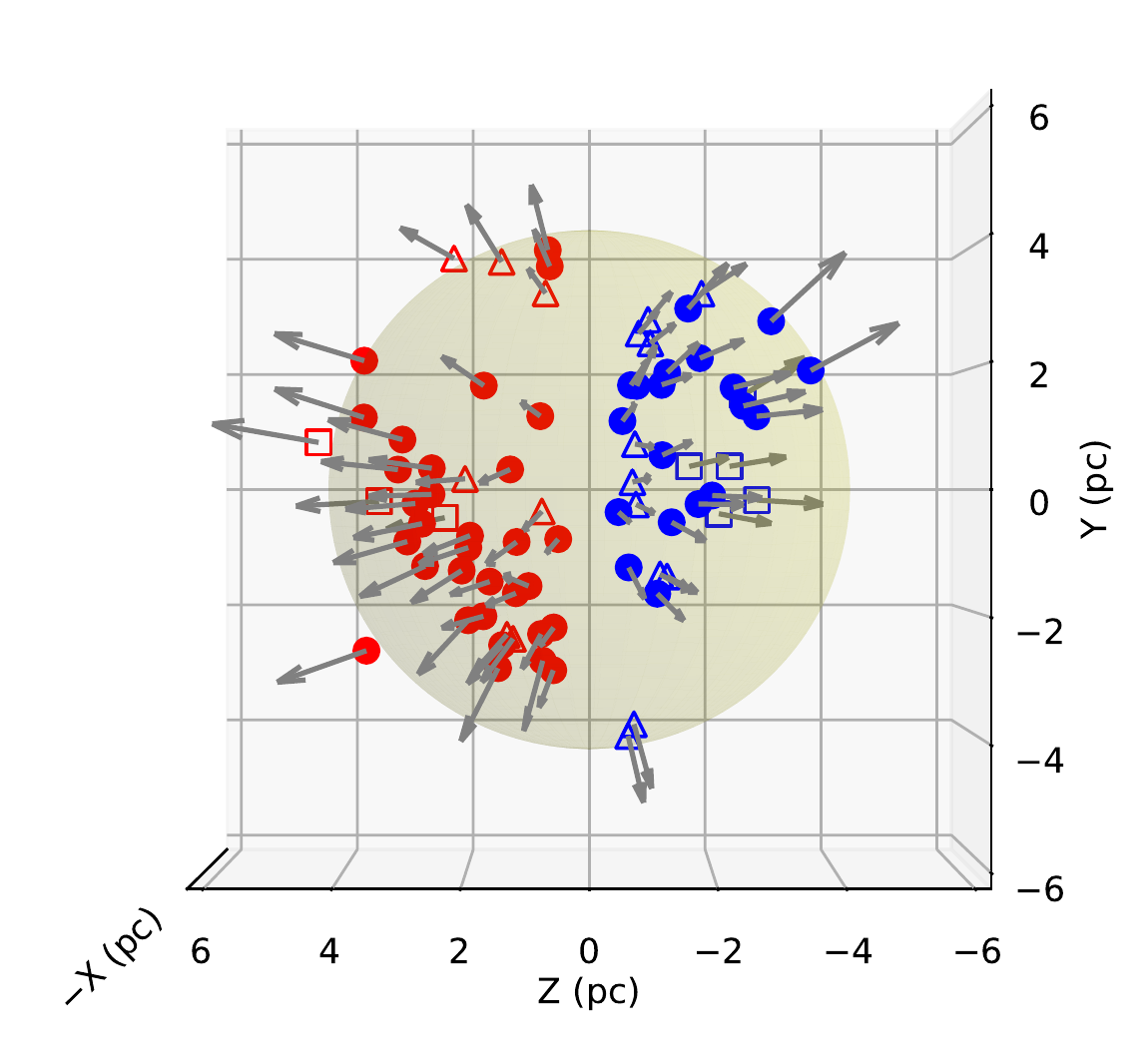}{0.49\textwidth}{(c)}
          \fig{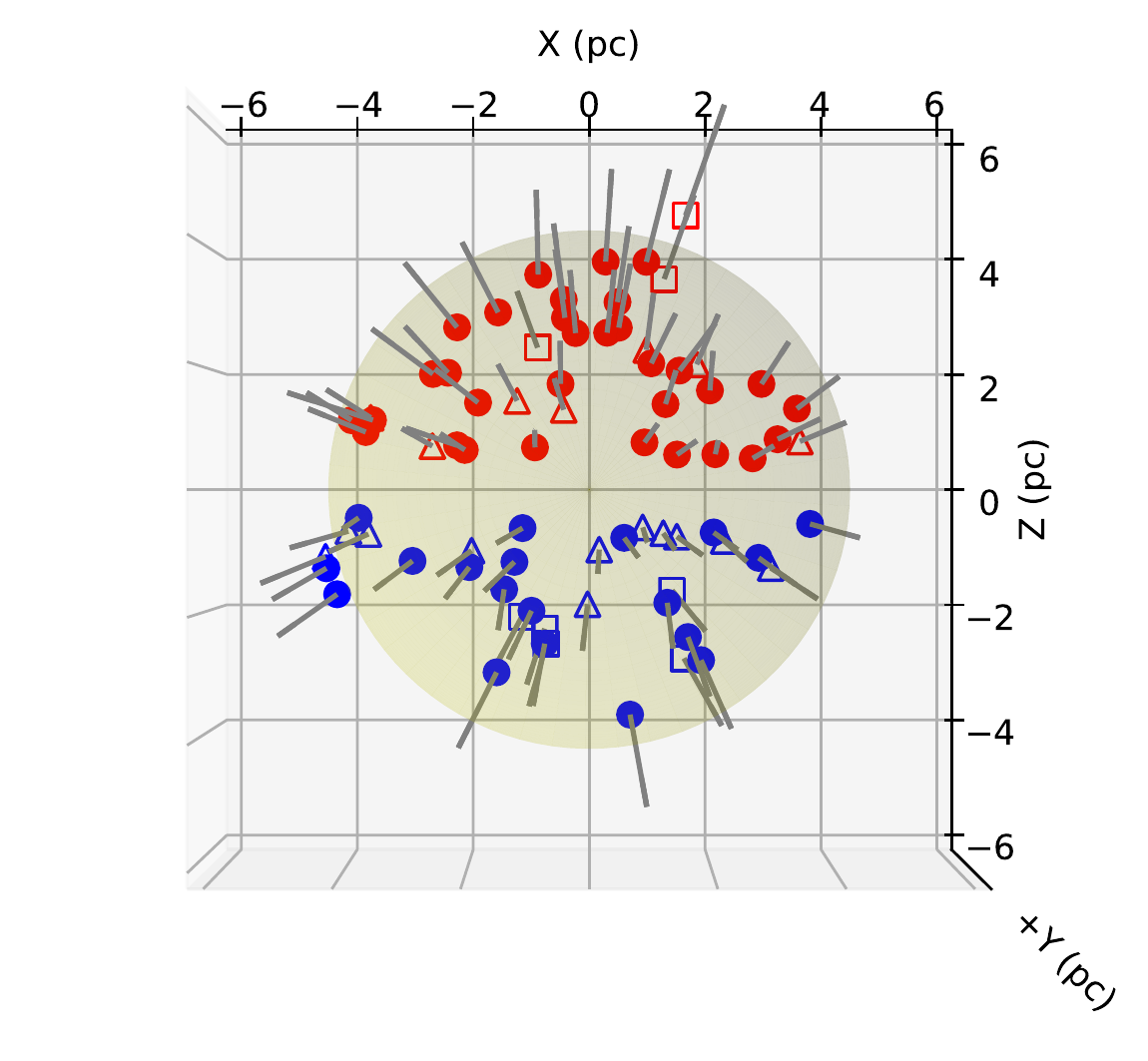}{0.49\textwidth}{(d)}
          }
\caption{3--D perspectives of the ejecta knots in Tycho. The red markers represent redshifted ejecta and blue are blueshifted ejecta. In (a) -- (c), we also overlay the ACIS measurements of ejecta velocities by \citetalias{sato17tych} and \citetalias{williams17}.   The circles, squares, and triangles show velocity measurements from our HETG sample, \citetalias{sato17tych}, and \citetalias{williams17}, respectively. For shared regions, we plot only our $v_r$ values. For the ACIS data, we include only those high-velocity regions with $v_r >$ 900 km s\textsuperscript{-1} (the ACIS systematic gain shift uncertainty). For knots where the proper motion was measured, the arrows point in the direction of the estimated 3D velocity. For the rest of the sample, the arrows point from the SNR center to the position of the knot. The length of the arrow represents the magnitude of the space velocity for each knot.  The pale shaded circle shows the approximate location of the main X-ray shell of Tycho. The axis along the line of sight increases into the page. In (a), the X and Y components represent the current locations based on each knot's R.A. and decl. In (b), the Z component of each knot is the measured $v_r$ multiplied by the age of the remnant. The Z component in (b) is likely underestimated due to the deceleration of the knots. In (c), the Z component from (b) is divided by the maximum forward shock expansion index, $\eta = 0.65$ \citep{katsuda10} to show a general approximation of the current physical positions of knots along the line of sight, accounting for their deceleration.  In (d), we show a ``top--down'' view of Tycho with the ejecta in the same positions as in (c). The Y-axis increases out of the page.
\label{fig:3d_view}}
\end{figure}

The X-ray emitting knots and filaments of the shocked ejecta gas in Tycho are distributed, in general, uniformly across the face of the SNR (Figures  \ref{fig:knotloc} and \ref{fig:3d_view}a). Our kinematic study of these ejecta knots shows that the overall spatial and velocity distributions of ejecta in Tycho are relatively smooth, in contrast to the case of Kepler's SNR where significant deviations from a spherical distribution, such as the ``Ears'' and nearly freely-expanding ejecta knots are present \citep{sato17kep,millard20}.  Our 3--D reconstruction of the ejecta distribution based on our radial velocity and proper motion measurements for a number of clumpy ejecta features indicates a relatively similar ejecta distribution between the eastern and western shells (Figures \ref{fig:3d_view}d). On the other hand, we find that the southern shell is dominated by redshifted ejecta (23 redshifted vs 6 blueshifted), while the majority of clumpy ejecta features in the northern shell are blueshifted (13 blueshifted vs 8 redshifted, see Figures \ref{fig:3d_view}b and  \ref{fig:3d_view}c).  The Chandra ACIS study by \citetalias{sato17tych} similarly revealed more blueshifted Si He-like and S He-like line-center energies in the north than in the south. The authors suggested that the observed discrepancy may be caused by a density enhancement of $\lesssim \sqrt{3}$ on the near side of the SNR compared with the far side. In this scenario, the density enhancement causes a stronger reverse shock on the northern near side.  Thus, more reverse--shocked ejecta is observed in the north than in the south.  A similar scenario could account for the north--south (N--S) differential in ejecta knots on the far side of the SNR. This N--S asymmetry of ejecta due to ambient density variation may be supported by the interacting density variations as reported by \citet{williams13} and \citet{katsuda10}.

%   some asymmetric ejecta structures in Tycho, based on our ejecta velocity measurements.

% 3D figure
%%%%%%%%%%%%%%%%%%%%%%%%%%%%%%%%%%%%%

% In 

% Reverse Shock
%%%%%%%%%%%%%%%%%%%%%%%%%%%%%%%%%%%%%%%%%%%%%%%%%%%%%%%%%%%%%%
\begin{figure*}[!h]
\centering
\gridline{
    \fig{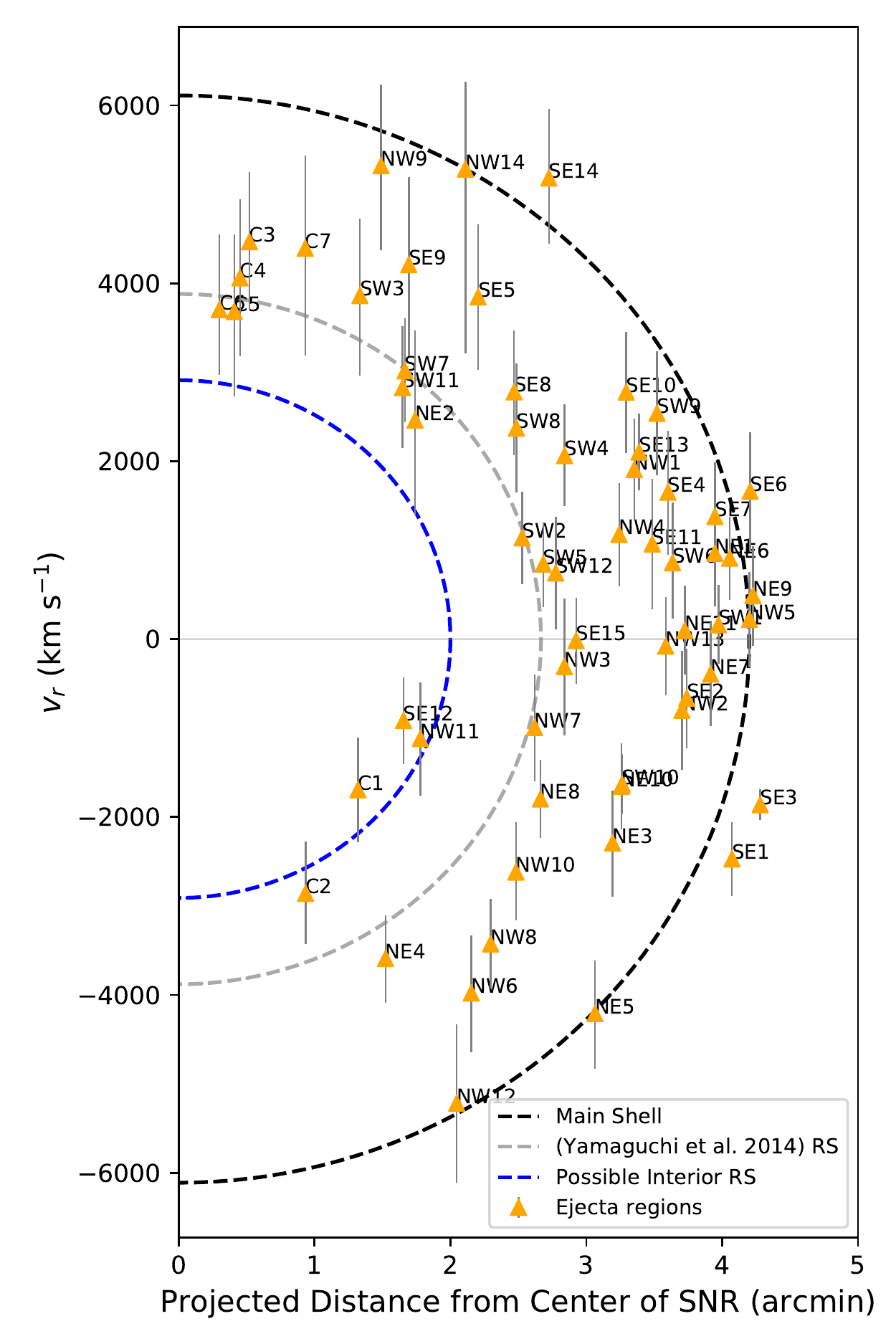}{0.32\textwidth}{(a)}
    \fig{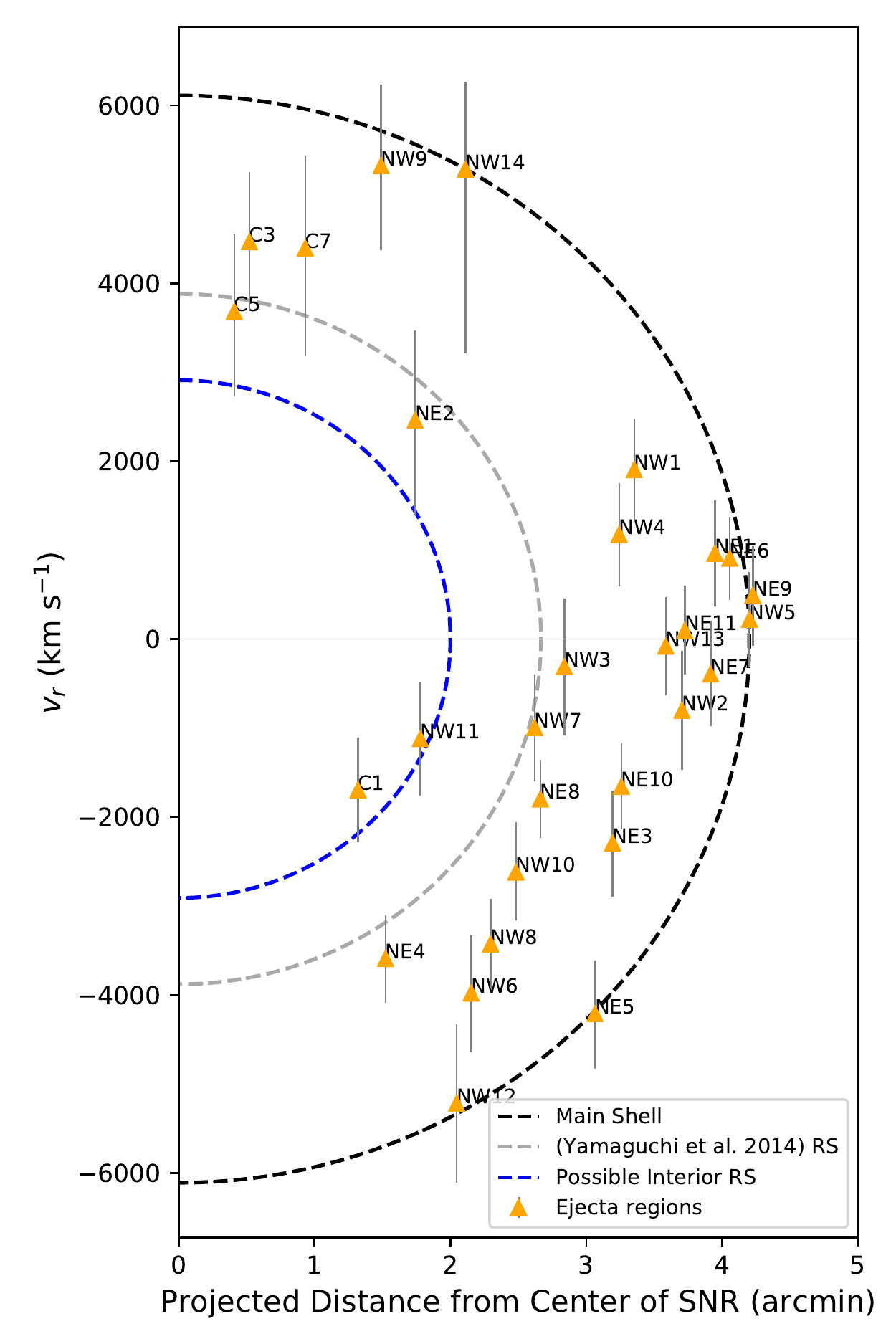}{0.32\textwidth}{(b)}
    \fig{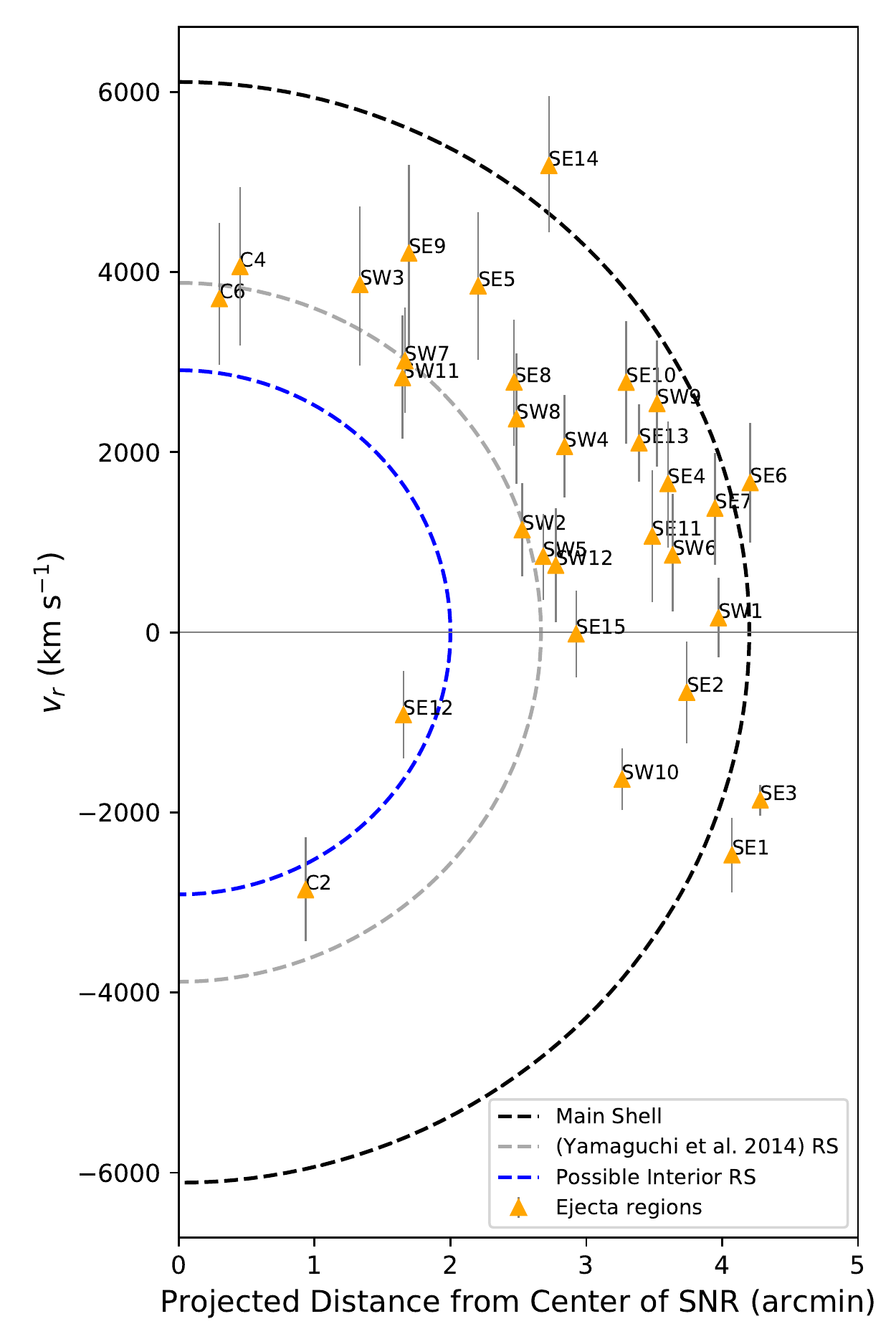}{0.32\textwidth}{(c)} }
\gridline{
    \fig{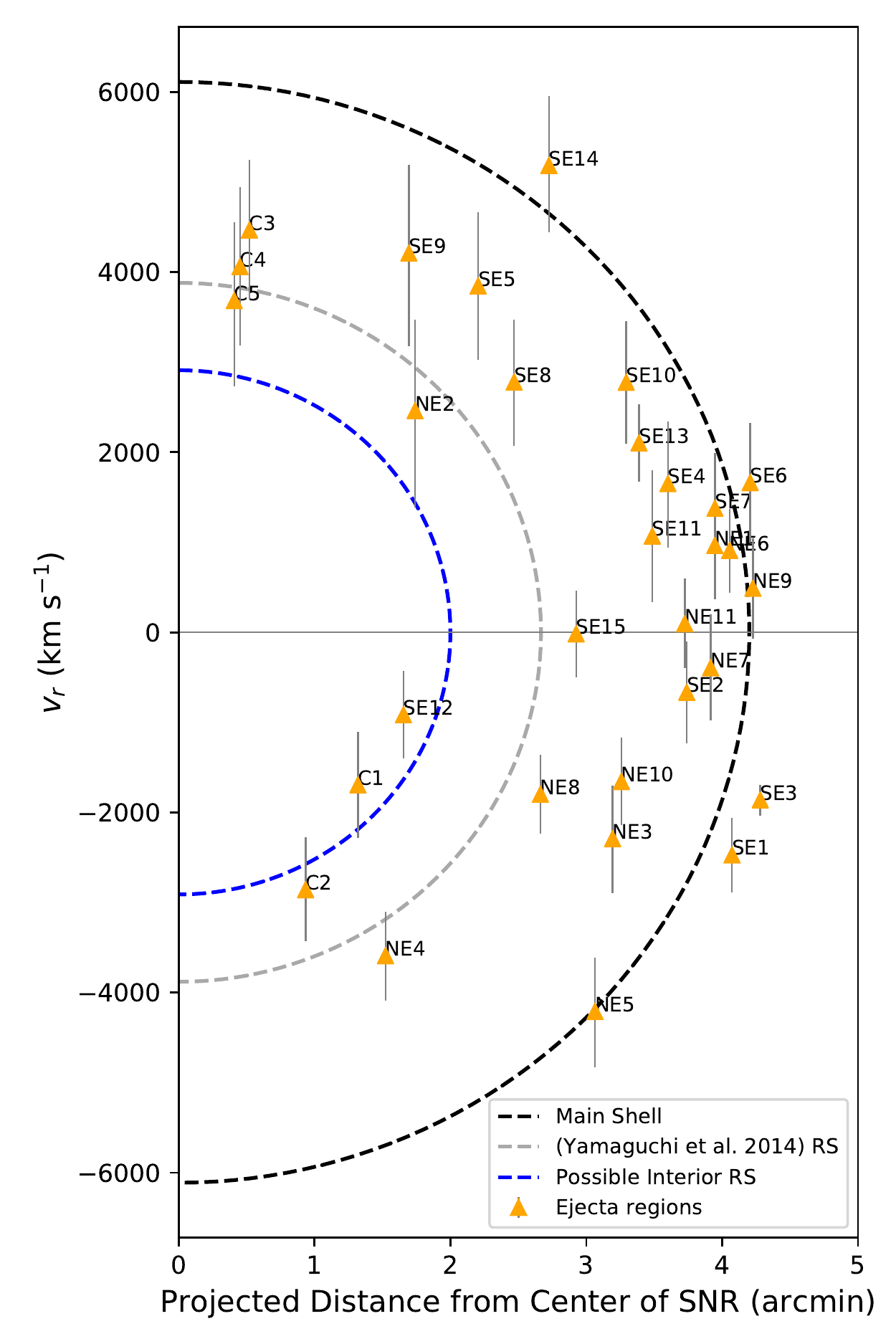}{0.32\textwidth}{(d)}
    \fig{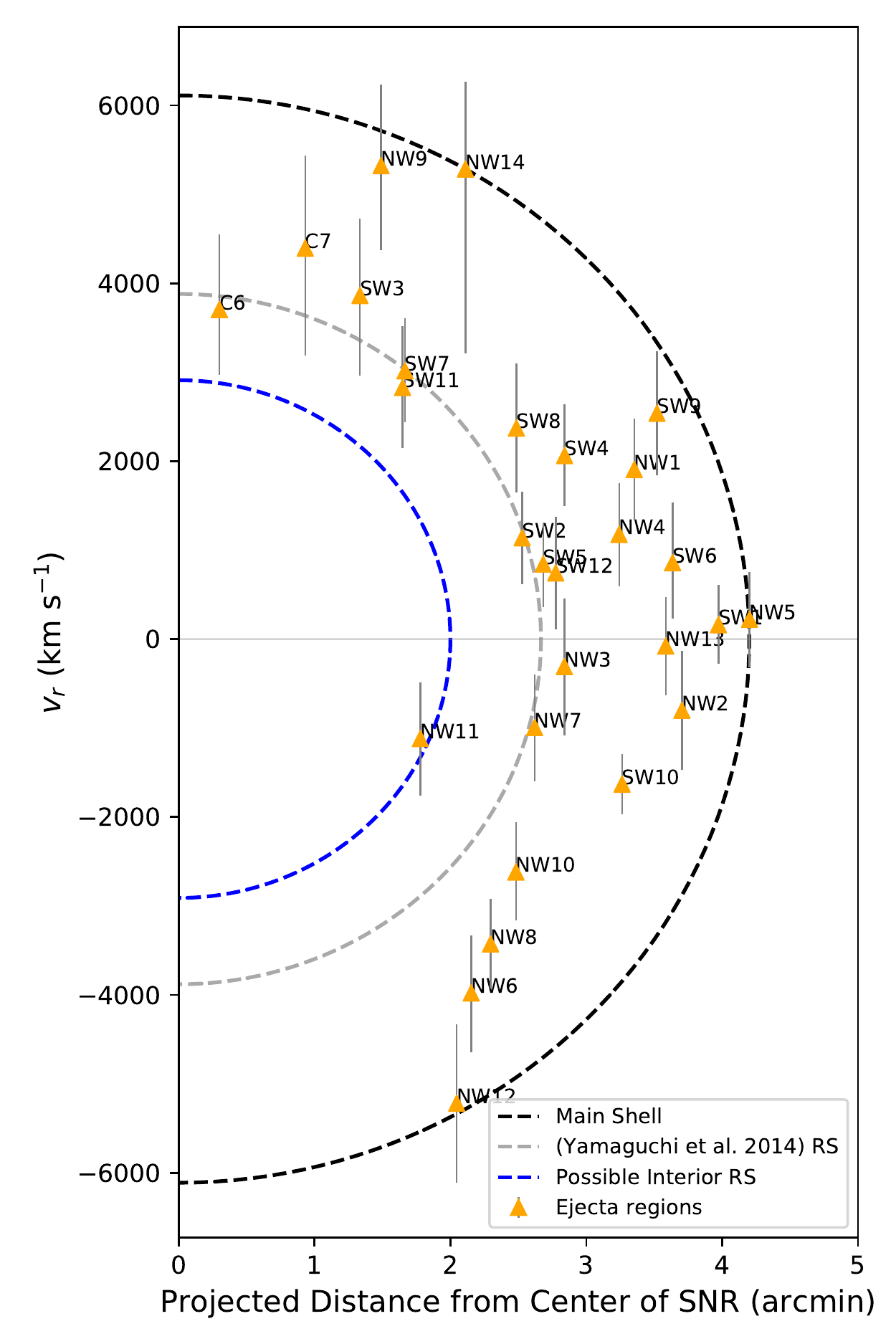}{0.32\textwidth}{(e)} }
%          \leftfig{vr_vs_r_plot_Tycho_d=3p5kpc_Warren2005_Our_Work_10-13-2021_show_NW_quadrant.pdf}{0.35\textwidth}{(d)}
%         }
%\gridline{\fig{KT_Eri-eps-converted-to.pdf}{0.3\textwidth}{(f)}}
\caption{In panel (a), the positions of ejecta knots in {\it v\textsubscript{r}} vs. {\it r} (projected angular distance from the center of the SNR) space. The black and gray dashed loci are the approximate locations of the outermost boundary of the main SNR shell and the reverse shock, respectively. The blue locus is a new potential reverse-shock location. A proportionality constant of 0\farcs{}041 (km s\textsuperscript{-1})\textsuperscript{-1} is applied to the loci based off the maximum expansion rate ($\sim$ 0.15 \% yr\textsuperscript{-1}) estimated by \citet{katsuda10}. Panels (b), (c), (d), and (e) show those knots located only in the northern, southern, eastern, and western hemispheres, respectively. \label{fig:rev_shock}}   
\end{figure*}

Although a variation in the ambient gas density surrounding Tycho is a plausible origin for the N--S ejecta differential, we may also consider that it could be due to an asymmetry in the early ejecta distribution immediately after the explosion.  \citet{seitenzahl13} simulated a range of Type Ia explosion scenarios and found that delayed detonation (DDT) models with fewer ignition points resulted in more asymmetric explosions. \citet{ferrand19} propagated a fully 3D N100 DDT model of \citet{seitenzahl13} into the SNR stage. They found that asymmetries in the explosion were required to explain the large-scale structures in X-ray maps of Tycho (specifically, the power spectrum of radius fluctuations around the rim, \citet{warren05}). \citet{ferrand21} further explored the early--stage evolutions of SNRs using the N5 (small number of ignition points) and N100 (large number of ignition points) DDT and pure deflagration models of \citet{seitenzahl13} and \citet{fink14}. The authors found that the N5ddt models produce a more asymmetric, dipolar remnant whose imprint lasts up to a few hundred years. It is not straightforward to directly compare the results of these simulations with the non-uniform ejecta distribution inferred from our velocity measurements. However, our work suggests the presence of an aspherical ejecta velocity distribution in Tycho.

\subsection{Explosion Center and Reverse Shock}\label{sec:rev_shoc_center}
We estimate the kinematic center of Tycho from our proper motion measurements of ejecta knots.   We choose knots which have both $\mu_{R.A.}$ and $\mu_{decl.}$ greater than the systematic uncertainty. We generally follow the technique employed in \citet{sato17kep}. Initially, we assume that each knot has moved at its current proper motion speed since the explosion (i.e., $\eta$ = 1) to estimate its 2-D starting position.   We average the starting positions of all individual knots to calculate the tentative ``initial'' kinematic center. Then, we calculate the new expansion index for each knot based on this tentative kinematic center, and trace its motion back to a new starting point, this time dividing the distance traveled by the expansion index to account for its decelerated motion.  We repeat this process until the average kinematic center converged on a single value (after about 25 iterations).  Our estimated kinematic center is R.A.(J2000) = 00\textsuperscript{h}25\textsuperscript{m}18\textsuperscript{s}.725 $\pm$ 1\textsuperscript{s}.157 and decl.(J2000) = +64$^{\circ}$08\arcmin 02\farcs5 $\pm$ 11\farcs{}2. This position is $\sim$ 13\arcsec{} southwest of the geometric center estimated by \citet{warren05}. The previously suggested candidates for the companion of Tycho's progenitor, Tycho G \citep{ruiz-lapuente04}, Tycho E \citep{ihara07}, and Tycho B \citep{kerzendorf13,kerzendorf18}, are located approximately 33\arcsec{} E,  14\arcsec{} NE, and 17\arcsec{} NW from our estimated center, respectively. Assuming a distance of 3.5 kpc, their transverse velocities since the explosion would be $\sim$ 1200 km s\textsuperscript{-1}, 500 km s\textsuperscript{-1}, and 600 km s\textsuperscript{-1}, respectively. This is in contrast to their recently measured proper motions \citep{kerzendorf13}, which would imply transverse velocities of 100 -- 200 km s\textsuperscript{-1}. However, since the positions of Tycho E and Tycho B are within a few arcseconds of the error ellipse of our estimated center (see Figure \ref{fig:pm_arrows}), they may have travelled a shorter distance if our kinematic center is representative of the explosion site. Thus, their transverse velocities since the explosion could be significantly slower, in line with the current values. Tycho G is located several arcseconds outside of the error ellipse, and therefore its current proper motion is still too low to account for the angular distance it would have travelled from our kinematic center since the explosion.

%0:25:18.7248
%+64:08:02.571

We plot our measured radial velocity for each knot against its angular distance from the center of the SNR in Figure \ref{fig:rev_shock}a. Figures \ref{fig:rev_shock}b and \ref{fig:rev_shock}c show the north--south asymmetry discussed in Section \ref{sec:3d_ejec_struc}, and Figures \ref{fig:rev_shock}d and \ref{fig:rev_shock}e show ejecta features projected in the eastern and western hemispheres, respectively.  We plot the main shell (the forward shock) and the reverse shock position from \citet{yamaguchi14} estimated from the location of Fe K$\beta$ emission generally in the NW quadrant of Tycho.  The bulk of the ejecta knots in our sample are positioned between the forward and reverse shocks, as expected. However, there are several knots positioned closer to the SNR center beyond the reverse shock. The locus of these inner ejecta knots appears to form a smaller reverse shock at $\sim$ 2.0\arcmin{} from the SNR center, or 75\% of the 2.6\arcmin{} radius for the reverse shock estimated by \citet{yamaguchi14}. According to models of dynamical evolution of SNRs \citep{truelove99} with an explosion energy of $1.2 \times 10^{51}$ ergs \citep{badenes06} and an ejected mass of 1.4 M\textsubscript{\(\odot\)}, ambient density variation by a factor of $\sim$ 4 (similar to that reported by \citet{williams13} for Tycho) may produce $\sim$ 30\% deeper-reaching reverse shock. These knots beyond the reach of the previously-known reverse shock are blueshifted, and therefore are positioned on the near side of the SNR. Thus, these inner ejecta knots may represent deviated parts of the reverse shock due to the shock interaction with denser medium on the near side of the SNR. Recently, X-ray proper motion measurements of Tycho's forward shock showed that its expansion has significantly decelerated from 2003 to 2015 (down to 40\% of its initial value, \citet{tanaka21}). The authors suggested that the forward shock may be encountering a non-uniform wall of dense gas, possibly created from the winds of the progenitor system. Our results may support the presence of a similar density variation along the line of sight.

%On 

%\subsection{Velocity Distribution of Ejecta} \label{subsec:vel_distribution}
  
%\newpage

\section{Conclusions} \label{sec:conclusions}

We have measured the radial velocities of 59 small ejecta features in Tycho's SNR using our deep 450 ks Chandra HETG observation. Based on these measurements, our 3--D reconstruction of Tycho shows a large-scale asymmetry where most knots in the northern half are blueshifted and thus on the near side, and most knots in the southern half are redshifted and therefore are located on the far side.  Ambient density variations across the near and far sides of the remnant might have caused non-uniformity in the formation of reverse shock, and thus resulted in the differences in the frequency of detected ejecta knots. Alternatively, the identified asymmetry could be caused by a non-spherical explosion of the progenitor.  

%Motivated by our observed asymmetric ejecta distribution, we also consider an intriguing post-explosion scenario of an ejecta shadow cast by the surviving companion.

%If 

For 37 of the 59 ejecta features in our sample, we measured their proper motions using archival Chandra ACIS data. We estimate an expansion center based on our measured proper motions. Combining the radial velocities and proper motions, we find space velocities up to 6000 km s\textsuperscript{-1}. The azimuthal distribution of our measured space velocities shows generally higher speeds of ejecta towards the SE. Regions with low velocity coincide with higher ambient density at the rim. However, the high-velocity SE regions do not coincide with comparatively lower ambient densities, and therefore probably resulted from higher kinetic energy being deposited in that direction from the explosion. Based on our detection of relatively lower radial velocities for several ejecta knots projected near the center of the SNR, we postulate a considerable ambient density variation along the line of sight (e.g., a higher density on the near side of Tycho).  

%The 

\bigskip
This work has been supported in part by NASA Chandra Grant GO7-18061X. TS was supported by the Japan Society for the Promotion of Science (JSPS) KAKENHI grant No. JP19K14749. JPH acknowledges support from NASA grant number NNX15AK71G to Rutgers University. We thank the anonymous referee for their useful recommendations that improved the explication of this work.

\end{document}